\documentclass[letter]{emulateapj}

\usepackage{amsmath,amsthm,amssymb}
\usepackage{lscape}
\usepackage{longtable}
\def\lsim{\mathrel{\rlap{\lower4pt\hbox{\hskip1pt$\sim$}}
    \raise1pt\hbox{$<$}}}                
\def\gsim{\mathrel{\rlap{\lower4pt\hbox{\hskip1pt$\sim$}}
    \raise1pt\hbox{$>$}}}                

\newcommand{\ergs}{{\mathrm{erg~s^{-1}}}}
\newcommand{\ergscm}{{\mathrm{erg~s^{-1}~cm^{-2}}}}
\newcommand{\ang}{{\mathrm{\AA}}}

\newcommand{\dd}{\mathrm{d}}

\newcommand{\lya}{Lyman~$\alpha$}

\newcommand{\llim}{L_\mathrm{lim}(\mathbf{\Omega},z)}
\newcommand{\mg}{\mathcal{M}(\mathbf{\Omega},z)}
\newcommand{\flim}{f_\mathrm{lim}(z)}
\newcommand{\slit}{T(\mathbf{\Omega})}
\newcommand{\sky}{\mathbf{\Omega}}

\shorttitle{A Keck Survey for Gravitationally-Lensed Lyman-$\alpha$ 
Emitters in at 8.5$\lsim$z$\lsim$10}

\shortauthors{Stark et al.}

\begin{document}

\title{A Keck Survey for Gravitationally-Lensed Lyman $\alpha$ Emitters 
in the Redshift Range 8.5$<$z$<$10.4: New Constraints on the Contribution 
of Low Luminosity Sources to Cosmic Reionization}


\author {Daniel P. Stark\altaffilmark{1}, 
Richard S. Ellis\altaffilmark{1}, 
Johan Richard \altaffilmark{1},
Jean-Paul Kneib \altaffilmark{1,2},
Graham P. Smith \altaffilmark{1,3},
Michael R. Santos \altaffilmark{4}
} 

\altaffiltext{1}{Department of Astrophysics, California Institute of Technology, 
MS 105-24, Pasadena, CA 91125; dps@astro.caltech.edu}
\altaffiltext{2}{Observatoire Astronomique de Marseille-Provence, 
Traverse du Siphon - BP 8, 13376 Marseille Cedex 12, France}
\altaffiltext{3}{School of Physics and Astronomy, University of Birmingham, Edgbaston, 
Birmingham, B15 2TT, UK}
\altaffiltext{4}{Space Telescope Science Institute, 3700 San Martin Drive,
Baltimore, MD 21218}
 
\begin{abstract}

We discuss new observational constraints on the abundance of faint 
high redshift Lyman-$\alpha$ emitters secured from a deep 
Keck near-infrared spectroscopic survey which utilizes the strong 
magnification provided by lensing galaxy clusters at intermediate redshift 
($z=0.2-0.5$). In each of 9 clusters we have undertaken a systematic `blind' 
search for line emission with NIRSPEC in the J-band within carefully-selected 
regions which offer very high magnifications ($\gsim\times$10-50) 
for background 
sources with redshifts $z\simeq 10$. The high magnification enables the 
detection of emission at unprecedented flux limits (10$^{41}$ - 10$^{42}$ 
erg s$^{-1}$),  much fainter than those of conventional narrow band imaging 
and other spectroscopic searches. As the comoving volumes probed are 
small, our survey is designed to address the important question of whether 
intrinsically feeble star forming galaxies could provide the dominant 
ionizing flux at $z\sim 10$. Our survey has yielded six promising 
($>5\sigma$) candidate Lyman-$\alpha$ emitters which lie between 
$z=8.7$ and $z=10.2$. We carefully discuss the validity of our detections
in the light of earlier, apparently false, claims. 
We then focus on the likelihood
that the detected line is Lyman-$\alpha$ from a distant lensed-source. 
All but one of the candidates remain undetected in deep HST optical images 
and lower redshift line interpretations can be excluded, with reasonable
assumptions, through the non-detection of secondary emission in 
further spectroscopy undertaken with LRIS and NIRSPEC.  We demonstrate
that rigorous confirmation of the Lyman-$\alpha$ hypothesis for all of our 
candidates will be a challenge with current facilities. 
Nonetheless, as a result
of our tests, we argue that at least two of our candidates are likely to be at 
$z\simeq 10$. Given the small survey volume, this suggests 
there is a large abundance of 
low luminosity star-forming sources occupying the era when the reionization 
process is thought, from independent evidence, to be ongoing. While the 
predicted reionization photon budget depends upon a large number of 
physical assumptions, our first glimpse at the $z\simeq 10$ Universe 
suggests that low luminosity star-forming galaxies contribute a 
significant proportion of the UV photons necessary for cosmic
reionization.

\end{abstract}


\section{Introduction}

After the epoch of recombination at z$\simeq $1000, neutral hydrogen
permeated the universe.  The earliest observable structure dates back 
to the beginning of this era at the surface of last scattering traced by the 
cosmic microwave background (CMB).  The next most distant observable 
represent the populations of galaxies, gamma-ray bursts, and quasars at 
$z\simeq 6.2-7.0$ \citep{Hu02,Iye,Totani,Kashikawa06,Fan06,Bou06b}.  
The interval between 
$z\simeq 1000$ and $z\simeq 6.5$ contains many landmark events:
the formation of the first stars, the assembly of first galaxies, 
the growth of 
the first supermassive black holes that power quasars, and the reionization 
of neutral hydrogen in the intergalactic medium. One of the primary driving 
forces in observational cosmology today is to illuminate the growth of these 
first structures in this important, but poorly understood, era.

Recently, important constraints have emerged on the extent of star formation 
activity that occurred in the latter part of this era. 
The WMAP temperature-polarization 
cross correlation signal on large angular scales \citep{Sp06} implies 
scattering of microwave photons by free electrons from ionizing sources at 
$z\simeq 10-12$, suggesting an early period of efficient star formation.  
Meanwhile, Hubble Space Telescope (HST) and Spitzer observations have 
uncovered a population of very massive (M$_{stellar}$ $\sim$ 10$^{10}$ - 
10$^{11}$ M$_\odot$) galaxies at z$\simeq$6 with well-established ($>$100 Myr) 
stellar populations \citep{Eg04,Ey05,M05,Y05,Stark06,Ey06,Y06}.  
The assembly of such massive galaxies by these redshifts 
requires significant star formation at yet earlier times. Observations 
of z-dropouts and J-dropouts with HST indicate a decline in the star 
formation activity of luminous Lyman-break galaxies out to $z\simeq 7-10$ 
\citep{Bou05,Bou06b}; if there is significant star formation 
activity in this redshift interval, these observations indicate 
it likely occurred in lower-luminosity systems.

Despite these indications of early activity, current facilities are 
poorly-equipped to undertake conventional searches for star-forming 
sources beyond $z\simeq 7$.  Even with HST and 8 meter class telescopes, 
continuum (`drop-out') and narrow-band imaging Lyman-$\alpha$ searches 
probe only the most luminous, rare, sources at these redshifts; the 
contribution to the reionization flux from these sources may well be small. 
While narrowband surveys have had significant success at detecting 
Lyman-$\alpha$ emitters at $z\simeq 6$ \citep{Hu98,Hu02,Hu04,MR04,Hu05,Stern05,
Kashikawa06,Shimasaku06} only one Lyman-$\alpha$ emitter 
has been convincingly detected at $z\simeq 7$ \citep{Iye}.  
At $z\simeq 9$, narrowband surveys for relatively luminous 
Lyman-$\alpha$ emitters ($\gsim 10^{42.5}$ 
erg s$^{-1}$) have yet to find any sources \citep{Willis05,Cuby06}.
While a few candidate Lyman-break galaxies beyond $z\simeq 7$ have been 
suggested from photometric data \citep{Bou05,Labbe06,Richard06, Bou06b, 
Henry07}, none has yet been spectroscopically confirmed, due largely to 
the difficulties inherent to observing in the near-infrared.

Recent observations of the faint end slope of the star-forming 
luminosity function at  $z\simeq 6$ \citep{yw04,Bou06} support the 
suggestion that the bulk of the integrated star formation at high redshift 
may arise from very low luminosity sources, as expected theoretically 
\citep{BL01,Wyithe06}.  However, forming stars in low mass halos is 
complicated by a number of feedback effects, any of which could significantly 
decrease the star formation efficiency. Clarifying whether low luminosity 
systems are abundant  at early times requires  observations probing well 
below the sensitivity limits obtainable with current facilities using 
conventional methods.

The study of lower luminosity sources rendered visible by the strong 
gravitational lensing induced by foreground clusters is the only current 
means to advance this quest before the next generation of telescopes is 
available.  Although a typical lensing cluster magnifies sources by 
$\times$5-10 over 1 arcmin$^2$ for sources at $z>7$,  faint signals can 
be boosted $>\times15$ on the so-called {\it critical line} - the location 
of the lensing caustic in the image plane. Spectroscopy and detailed 
studies become feasible at otherwise impossible (unlensed) limits. For 
several objects recently located via this technique 
\citep{Ellis01,Kneib04,Santos04,Eg04}, stellar continuum slopes, 
Lyman-$\alpha$ profiles and star formation rates have been determined 
for sources whose intrinsic flux is close to the faintest direct imaging 
limits so far reached, $I_{AB}\simeq 30$ \citep{Beckwith06}. Until the 
availability of more powerful ground and space-based facilities, scanning 
the critical lines of massive clusters provides possibly the only 
reliable means of estimating the abundance of low luminosity star forming 
sources in the 7$<z<$12 interval over which cosmic reionization is thought 
to take place.  

This paper presents the results of a spectroscopic survey for 
strongly-lensed Lyman-$\alpha$ emitters at 8.5$<$z$<$10.4 using the 
NIRSPEC near-infrared spectrograph on the Keck II telescope.  The 
survey follows logically from a very similar, successful one undertaken 
at $z\simeq 4-6$ using the LRIS optical spectrograph on the Keck I
telescope \citep{Santos04} to which the reader is referred. 

A plan of the paper follows.  We contrast our lensing search technique 
with other, more conventional probes of early star forming sources in $\S$2. 
The NIRSPEC observations and reduction methods are presented in $\S3$.  
Here we describe the candidate selection criteria and derive the survey 
sensitivity function. A set of 6 promising candidate $z\simeq 10$ 
Lyman-$\alpha$ emitters are presented in $\S$4.  Here we discuss in 
detail both the reality of the line detections and the likelihood 
that the line identifications represent high redshift Lyman-$\alpha$. 
Noting the likelihood that a significant fraction of our sources may 
lie at $z\simeq 10$,  we infer the abundance of high redshift low 
luminosity sources and compare these with other constraints in 
$\S5$.  In $\S$6 we discuss whether the abundance so derived might represent
a significant fraction of that necessary to complete cosmic reionization. 
We present our conclusions in $\S$7.

Throughout the paper, we have assumed a flat universe and 
($\Omega_m,\Omega_\Lambda$)=(0.27,0.73) following the results presented in 
the initial WMAP data release \citep{Sp03}.

\section{Gravitational Lensing and Survey Strategies - A Critique}

Strong gravitational lensing by intermediate redshift  galaxy clusters offers
the capability of extending the current generation of high redshift galaxy 
surveys so that lower luminosity sources can be brought into view.  
This added boost in sensitivity is particularly effective at the highest source 
redshifts ($z\gsim 7$) where the bulk of star formation activity may be taking 
place in low mass dark matter halos \citep{BL01}.

Lensing increases the apparent area of a background source while conserving 
the surface brightness; thus the gain in sensitivity is the magnification 
factor, $\mathcal{M}$.  Rich clusters of galaxies at intermediate redshifts 
(e.g. $z\simeq 0.2-0.5$)  amplify high redshift sources by factors of 
typically $\mathcal{M}\simeq 3-5$ over areas of a few arcmin$^2$. However, 
boosts of $\mathcal{M}\simeq 10-50$ are possible near the {\it critical line} 
- the theoretical locus of points in the image plane corresponding to the 
caustic of  infinite magnification.  Clusters are usually characterized 
by both an inner and an outer critical line, the location of which varies 
as a function of source (background) redshift.  

The accurate location on the sky of these critical lines can only be 
determined in the case of clusters for which detailed mass models are 
available.  Such a 
mass model  is primarily constrained by the positions and redshifts of lensed 
features \citep{Kneib96}.  A mass model can be validated, for example, 
by securing spectrosopy of lensed features whose redshifts were predicted 
by the mass model using an `inversion' technique \citep{Ebbels98}. 
Currently, such high quality mass models are only available for $\simeq$20 
clusters; however this should increase in the future with additional 
surveys \citep{Ebeling01}.

However, the gain in sensitivity afforded by strong lensing comes with a 
loss in survey area, compared to an equivalent unlensed survey, by a factor 
of $\mathcal{M}$. This is an important point as the area of high magnification
(say $\mathcal{M}>$10) is already quite small ($\simeq 0.1$ arcmin$^2$). 
Provided, as seems reasonable ($\S$1), that the slope of the faint-end 
slope of the luminosity function is very steep, then strong lensing surveys 
should be a very effective way to characterize the amount of star formation 
at $z>7$. 

Observers are considering several ways to conduct surveys for 
$z>$7 Lyman-$\alpha$ emitters. Before describing the present survey, 
we therefore consider it instructive to 
examine the benefits and drawbacks of the different strategies.  

The primary survey techniques used thus far are narrowband imaging 
\citep{Hu02,Hu04} and spectroscopy \citep{Martin06}. Neither 
technique has yet been routinely used in the near-infrared spectral region 
appropriate for $z>$7 and both methods can, in principle, be combined 
with strong lensing by focusing on fields containing rich clusters. At 
optical wavelengths, unlensed narrow-band imaging techniques typically 
reach flux limits of $2\times$10$^{-17}$ ergs cm$^{-2}$ s$^{-1}$, 
whereas spectroscopic techniques probe to $5\times$10$^{-18}$ ergs cm$^{-2}$ 
s$^{-1}$ \citep{Santos04}. 

In the near-IR, a longslit spectrograph such as NIRSPEC offers a 
$\sim$8 \AA~ resolution over a moderate redshift interval ($\Delta z$=2). 
Although bright atmospheric emission lines restrict visibility for 50\% of the 
redshift range in the J-band, for a resolving power R$\simeq $1500, the gain 
in redshift coverage is still a factor of 10-100 over the narrow band 
technique ($\Delta z$=0.01-0.1). Per exposure, there is however a very 
small field of view ($\simeq $30 arcsec$^2$) c.f. narrowband imaging  
($\simeq 10$ arcmin$^2$ for a cluster but much larger for panoramic field 
surveys). 

Assuming the characteristics for current instrumentation and their detectors, 
we compare the efficiencies of the two methods in the lensing case in 
Figure 1. Details of the computation are given in the Appendix.   Both 
lensed survey techniques probe at least an order of magnitude deeper than 
conventional near-IR Lyman-$\alpha$ narrowband surveys.  Of the two lensing 
methods, long-slit spectroscopy is better suited for detecting the faintest 
Lyman-$\alpha$ emitters for detectors 1024$\times$1024 pixels in size.  This 
can be explained as follows.  While a narrowband imager covers a larger 
area on the sky, only a very small fraction of this area is highly magnified 
(e.g. $>$10$\times$).  The small gain in the highly-magnified survey area 
is insufficient to compensate for the significantly smaller redshift coverage. 
Moreover, the $\simeq $100 \AA~ bandwidth typical for many narrowband 
filters results in significantly poorer sensitivity limits.  Taking the 
various factors into consideration, a long slit spectroscopic survey 
represents the more efficient technique for detecting sources with 
star formation rates below 10$^{42}$ erg s$^{-1}$, and hence is the optimal 
strategy for determining whether reionization was predominantly caused by 
large numbers of feeble sources.  

\section{NIRSPEC Critical Line Survey}

\subsection{Cluster Sample}

Our goal is to constrain the abundance of sources up to an order of 
magnitude below current survey sensitivity limits. Accordingly, following
the discussion above, we have conducted a spectroscopic survey for
Lyman-$\alpha$ emission along the critical lines of 9 clusters at 
$z\simeq 0.2-0.5$.
We limited our observations to those clusters with mass models 
that are well-constrained by the combination of HST imaging and  
ground-based spectroscopy of multiply imaged systems. The list
of survey clusters and the associated mass models are summarized 
in Table 1. 

Many criteria entered into the final selection of our 9 clusters. An
essential criteria was the availability of deep HST imaging for constructing 
samples of multiply-imaged systems \citep{Kneib96, Broadhurst05a}. The most 
well-understood clusters in our sample contain 10-30 
multiply-imaged systems, a signification fraction of which have spectroscopic 
redshift (e.g. Abell 1689 has 33 such systems of which 21 are 
spectroscopically-confirmed, Richard  et al, in prep). This data is
necessary to astrometrically pinpoint the location of the critical line
along which the NIRSPEC slit is oriented.

Deep HST optical imaging is also essential for verifying, via the absence 
of any detectable continuum, that any lensed emitters are genuinely at 
high redshift. Due to neutral hydrogen absorption we would not expect
any of our $z\simeq 10$ emitters to be seen at optical wavelengths and
HST provides the deepest verification.

A final consideration is the availability of deep optical {\it spectroscopy}
which is helpful in order to check for associated emission lines that
might arise if any NIRSPEC line detection arises from a lower redshift
source. Given the comprehensive critical line survey undertaken by
\citep{Santos04} using LRIS, where possible we chose the same 
clusters and explored the same critical line regions.

In practice, telescope scheduling and weather losses compromised
a perfect adoption of the latter criteria. However, for each of the 9 
clusters, the mass model, the critical line location and associated 
magnification properties are well understood.

\subsection{Observations and Data Reduction}

We utilized the Near Infrared Spectrometer (NIRSPEC, \citealt{Mclean98})
mounted on the Keck II 10 m telescope at Mauna Kea to perform our survey.  
Observations were performed in the J-band (1.143-1.375 $\mu$m) with 
a slit 42\arcsec\ long and 0\farcs76 wide.  We used the low-resolution 
mode of NIRSPEC with the 75 line mm$^{-1}$ grating which offers a 
net resolving power of R$\simeq $1500 and a spectral 
resolution of $\simeq $8 $\ang$.  At the wavelengths sampled within 
the NIRSPEC J-band filter, any detected Lyman-$\alpha$ emission would 
correspond to sources with redshifts between $z=8.5$ and $z=10.4$. 

Figure 2 shows how the targeted slit positions match the predicted
location of the $z\simeq $9 critical  lines for each of the 9 clusters in 
our survey. To maximize efficiency, where a choice was available we 
selected regions where the critical line is straight enough to provide 
high magnification across the entire slit.
  
Typically, 2-5 slit positions were observed per cluster. For each slit 
position, we took 6-10 ten minute exposures using a three point 
dither pattern in which the telescope was offset $\pm$3\arcsec\ 
along the slit.  Table 2 summarizes the details of our NIRSPEC 
observations.  In four observing runs, we observed 35 slit positions 
corresponding to a total sky area of 0.3 arcmin$^2$.

Spectra were flat-fielded and sky-subtracted using IDL scripts 
written by G. Becker (2004, private communication).  Following
techniques described in \cite{Kelson03}, the two-dimensional
spectra were {\it not} straightened prior to sky subtraction, thus
ensuring sub-pixel sampling of atmospheric emission lines
which, in turn, led to a significantly improved removal of the 
sky background. 

The camera distortion and spectral curvature were computed
by fitting standard star traces along the slit  and atmospheric emission 
lines across the dispersion axis.  Knowledge of the distortion and 
spectral curvature was used to construct arrays that provide proxies 
to the slit position and wavelength for each exposed pixel on the 
NIRSPEC detector. Using this information, sky background modeling 
was performed using a two-dimensional b-spline fit, where a 
low-order polynomial is fit to the slit illumination and a b-spline is 
fit along the dispersion axis.  The wavelength calibration is 
subsequently computed using atmospheric emission lines.  
For each exposure, we also compute 2D-variance arrays by 
summing the contribution from the dark current, flat-field, read noise, 
sky background, and source counts to the variance.  

Following techniques introduced by Santos et al (2004), 
we determined the astrometric position of each spectral exposure by 
registering images from the NIRSPEC slit-viewing camera 
(SCAM) to HST images of the same field. WCS coordinates 
were computed for each exposed pixel on the detector using the
slit position grid described above.  Since these are galaxy 
cluster fields, many bright objects appear in the 46$\times$46
arcsec$^2$ field of view, enabling accurate registration on to 
the WCS of the HST images with a typical rms of 0\farcs15 
(less than 25\% of the slit width).  Offsets between exposures of a 
given slit position are calculated from the registered SCAM images, 
and the 2D-spectra and their associated 2D-variance arrays are 
subsequently shifted and combined. To remove cosmic rays and bad pixels, 
we median combine the data rejecting the brightest and faintest frame at each 
pixel.  

Our ability to detect faint emission lines is strongly dependent upon 
the accuracy with which the offsets between the different exposures 
of a slit position are known.  If a bright emission line from a foreground 
galaxy lies serendipitously on the slit, the accuracy of the offsets can 
be determined by comparing the shifts derived from the SCAM images 
to the offsets between the centroid of the emission line.  Suitable emission 
lines are present in several of our slit positions and applying the 
aforementioned 
test, we find the offsets determined from the SCAM images are typically 
good to 0\farcs1.  At  less than 15\% of the slit width, this uncertainty 
does not compromise the S/N of faint emission features.

One additional concern in dealing with faint emission line objects is 
the possible transverse drift of the object across the slit from 
frame to frame when the seeing is less than a slit width \citep{Pello04,Weatherley04}.  
This would cause the centroid of the emission 
line to drift in the {\it dispersion} direction of the spectrun.  
Uncorrected, this 
would increase the area over which the emission line flux is spread, thereby 
reducing the observed S/N of the emission line in the final stacked 
spectrum.  We quantify the magnitude of the transverse drift by measuring 
the frame-to-frame centroid of a very bright [OIII] emission line in NIRSPEC 
observations taken with exceptional seeing (0\farcs45).  We find that the 
standard  deviation in the centroid position over 9 frames is 0.8 pixels, or 0\farcs2.
Adding this measured tranverse drift in quadrature to the seeing FWHM 
only increases the FWHM of the line by 5\%.  Given that the seeing 
was never better than the value quoted above, it is clear that the 
drift has a negligible effect on the detectability of emission lines. 

\subsection{Survey Sensitivity}

We now determine our survey sensitivity function.  We first discuss 
the limiting line flux as a function of wavelength and source redshift.  
Converting this flux to a limiting source luminosity requires knowledge of the 
magnification across the NIRSPEC slit.  In practice, we will quote our
limit in terms of that line flux we can expect to detect, at $5\sigma$
confidence, for an unresolved emitter whose rest-frame line width is 
typical of a faint, but very well-studied, $z\simeq$5.7 system 
\citep{Ellis01}.

To accomplish this, we computed the total noise as a function of 
dispersion and slit position using the variance arrays obtained in the 
data reduction ($\S$3.1). This comprises the sum of the variance from 
the sky background, read noise, flat-fielding, dark count as well as
the source counts. Flux calibration of the variance arrays was performed 
using observations of spectroscopic standard stars.  The 5$\sigma$ 
limiting Lyman-$\alpha$ line flux was then computed as a function of slit 
position and wavelength by calculating the signal that is 5 times the 
root of the variance in an aperture whose spatial dimension is twice 
the measured seeing disk (typically 0\farcs7) and whose spectral 
dimension equals the Lyman-$\alpha$ line width, $\sim$300 km s$^{-1}$ of
the lensed system studied with high precision by \cite{Ellis01}. 

We find the median 5$\sigma$ limiting line flux across a typical slit position 
with 1.5 hours of integration is $2\times$ 10$^{-17}$ erg 
cm$^{-2}$ s$^{-1}$ (see Figure 3); the sensitivity varies between 0.9-3
$\times$ 10$^{-17}$ erg cm$^{-2}$ s$^{-1}$ depending on the 
proximity to atmospheric OH lines. This flux limit is similar  
to that reached in the $z=8.8$ narrowband survey of GOODS-S discussed in 
\cite{Cuby06} (1.3$\times10^{-17}$ erg cm$^{-2}$ s$^{-1}$) and 
slightly less sensitive than the $z=8.8$ narrowband survey 
presented in \cite{Willis05} (3.3$\times10^{-18}$ erg cm$^{-2}$ s$^{-1}$);
however the lensing magnification enables the detection of sources 
that are significantly less luminous than those detectable in 
either conventional narrowband survey.

\subsection{Cluster Magnification}

As discussed in our precursor optical paper \citep{Santos04},  the 
cluster mass models listed in Table 1 provide a 2-D map 
of magnifications for any given source redshift. The details by which  
these magnifications are realized follow techniques described in
detail by \citep{Kneib96}.  Further discussion of the code, LENSTOOL,
used to create the mass models for systems for which rich detail is
available Jullo et al, ({\em in prep} and \cite{Smith05}. From each mass 
model, the magnification can be computed as a function of  source redshift and 
position.  

In practice, for each position and wavelength, the mass models produce 
a matrix of the convergence and shear. The magnification is then determined 
from the convergence $\kappa$ and shear $\gamma$ using the following 
equation:

\begin{equation}
\mg = \frac{1}{[1-\kappa(\sky,z)]^2 -\gamma(\sky,z)^2} 
\end{equation}
where $\sky$ is the position on the sky and $z$ is the source redshift.

A key question that often arises in the consideration of critical line 
surveys is the
accuracy with which the location of the line can be pinpointed and
also that of the associated magnification map. To address this 
question, we have used a Markov chain Monte Carlo sampling algorithm 
within LENSTOOL (Jullo et al., {in prep}) to map the probability density 
in the 
mass model parameter space for each cluster. We then generated a magnification
map for each model and in turn computed the mode of the distribution at 
each pixel as well as the range of 
magnification factors which bracket the mode by $\pm$34\%. 
This we take as the uncertainty in the magnification.  The  
errors derived for each cluster are applied in the determination 
of both the candidate luminosities ($\S5$) and the survey volume ($\S$6.1).

Taking each exposed pixel on the NIRSPEC detector as an element of 
solid angle, the convergence and shear is computed as a function of 
position via interpolation between the NIRSPEC coordinate grid and 
that of the convergence and shear maps.  For each NIRSPEC pixel, the 
associated redshift at which a Lyman-$\alpha$ emitter would be located is 
calculated using the wavelength solution described above. The convergence, 
shear and Lyman-$\alpha$ redshift is then inserted into equation 1 and the 
magnification $\mathcal{M}$ at each pixel determined.  Likewise, 
the magnification uncertainty map is interpolated onto the NIRSPEC 
coordinate grid enabling the computation of the error as a 
function of slit position.  Both the magnification and associated 
error maps were smoothed using a  1\farcs0 median box filter 
to ensure that our results are not corrupted by sharp high peaks in 
the magnification distribution.  The change in the magnification 
error with redshift is negligible; hence we adopt the uncertainity at 
$z = 9$ to be the magnification error for all Lyman-$\alpha$ redshifts 
probed by our survey.

As an example of the magnifications sampled, the wavelength-independent 
magnification along one of our survey slit positions is displayed in 
Figure 4 for three different redshifts. The magnification is $>\times$20 
over nearly the entire slit and fairly constant with redshift.  
We also plot the associated 1$\sigma$ uncertainty as a 
function of position along the slit at one of the redshifts.  Even 
allowing for the uncertainty 
in the lensing model, the magnification is still expected to be uniformly 
high across the entire slit.  This is largely a result of the fact that we 
only select clusters with well-defined mass models. Furthermore, in regions 
where the location of the critical line is known less precisely, there is 
generally a region $\simeq$2\farcs~ in width around the critical line where 
the magnification factor is greater than fifteen, so the positioning of the 
NIRSPEC slit is always in a region of high magnification.

Figure 5 presents a histogram of the magnification in our survey.  Throughout 
the entire survey area, the median magnification provided to background 
sources is $\simeq\times$20, a factor of two larger than in the optical survey 
conducted by \cite{Santos04}. The increased magnification is 
due to several factors.  First, ultra-high magnification regions 
($\times$15-30 throughout the slit-area) were preferably selected to 
allow very low-luminosity systems ($\lsim 10^{42} erg s^{-1}$) to be probed. 
Second, since the NIRSPEC longslit covers a factor of three smaller angular 
extent on the sky than the LRIS longslit used in \cite{Santos04}, we were 
able to avoid regions where the critical line curved off the slit, resulting
in high magnification uniformly across the slit. 

\subsection{Limiting Lyman-$\alpha$ Luminosity}

Given our knowledge of the limiting Lyman-$\alpha$ flux (\S3.2) and the 
spatially-dependent magnification (\S3.3), we are now in a position
to calculate the limiting source Lyman-$\alpha$ luminosity, $\llim$. This
we define as the least luminous source detectable (with 5$\sigma$
confidence) in a given volume element, allowing for the cluster 
magnification $\mathcal{M}$. 

Allowing each NIRSPEC pixel to correspond to a volume element,
we can calculate the limiting luminosity across each slit in the survey
according to the following:

\begin{equation}
\llim = \frac{4 \pi (1+z)^2 D_\mathrm{c}^2(z)}{\slit} \frac{\flim}{\mg}.
\end{equation}
where D$_\mathrm{c}$(z) is the comoving distance
of the volume element and $\slit$ is the slit transmission.  

In the absence of the lensing magnification, the typical limiting 
Lyman-$\alpha$ 
luminosity is 2$\times$ 10$^{43} \ergs$ in between atmospheric OH lines. 
However, more than 50\% of the survey area is magnified by $ > \times 30$, 
enabling the detectability of sources as faint as 7$\times$ 10$^{41} \ergs$  
over $\simeq $0.15 arcmin$^2$ in the image plane.  Conventional 
narrowband Lyman-$\alpha$ surveys at $z>6$ typically reach limiting 
luminosities
between 10$^{42}$ and 10$^{43}$ erg s$^{-1}$. Our NIRSPEC survey 
probes over an order of magnitude fainter and thus can clearly provide 
constraints on star forming sources that are otherwise out of reach of
traditional methods, at least with current facilities.

\section{Identification and Reality of the Candidate Lyman-$\alpha$ Emitters}

At the faint limits now being probed, we have found the reliable identification
and verification of distant  Lyman-$\alpha$ emitters to be a very challenging 
endeavor, 
even with the most powerful facilities available to us.  The burden of
proof that a detected line is (a) real and (b) truly arises from a 
highly-redshifted Lyman-$\alpha$ emission line is very great given earlier 
controversies \citep{Pello04,Bremer04,Smith06}.  So far as the 
identification process is concerned, the key issue is to ensure
candidate emission features are not spurious and do not arise from 
detector artefacts \cite{Weatherley04}. Concerning the redshift 
verification, a new set of
challenges emerge at $z\simeq $10 since additional (confirmatory) emission
lines are not available from ground-based facilities (see \S5).

Beginning with the identification process, each of the sky-subtracted 2-D
spectra were independently inspected for Lyman-$\alpha$ candidates by three of 
the authors (D.P.S,  R.S.E, and J.R.).  We demanded that any candidate feature 
be both extended over multiple pixels (in order to differentiate a signal from 
bad pixels and other detector artifacts) and located away from contaminating 
OH airglow lines. The three lists of line candidates each contained $\simeq$ 25 
possible features with many being in common between two of the three lists.  
A final catalog of six promising candidates was constructed by selecting 
those features with the highest significance common to all three observers.
These are displayed in Figure 6.  The S/N ratios of these candidate emission 
lines, defined as in \S3.2, varies between 5 and 8 (see Table 3).

A spurious feature could arise if the reduction process does 
not adequately remove a cosmic ray or bad pixel.  Obviously dithering
the telescope so the candidates move up and down the slit eliminates
most detector anomalies. However artefacts can remain if the spectra 
are rebinned during the wavelength calibration; in this case, such bad pixels 
would be smoothed out and appear similar to emission lines 
\citep{Weatherley04}.  Fortunately, we do not face this problem in our 
Keck dataset since our reduction technique altogether avoids rebinning the 
spectra (see $\S$3).  Nevertheless, one may still worry that the measured 
emission line flux originates from one or two frames with overlapping 
cosmic rays or bad pixels.  We have tested this possibility by median 
combination of all of the exposures of slit positions containing our 
candidates and rejecting the three highest pixels.  We also median combined 
half of the exposures (chosen randomly) for each candidate.  In both tests, 
all six candidates are still clearly visible (albeit at lower S/N for the 
latter test), suggesting that our emission features do not arise from flux 
in only one or two frames.  

Finally, it is conceivable that the noise distribution of the spectral data is
such that $5\sigma$ noise features are much more common than would be 
expected under Gaussian statistics.  Such a curious situation might arise if
there was some unforeseen property of the NIRSPEC detector. In this case 
our candidate emission features might simply be peaks in the noise distribution.  
We investigated this possibility by computing the flux in apertures 
centered at random positions in areas of the spectra where the
sky-subtraction is clean.  The aperture size was matched to the expected 
size of a Lyman-$\alpha$ feature at high redshift: the spatial FWHM was taken 
to be the size of a typical seeing disk, 0\farcs6 and spectral FWHM 
corresponding to typical line width of Lyman-$\alpha$ emitters, 300 
km s$^{-1}$) 

A normalized histogram of the distribution of summed 
aperture fluxes is presented in Figure 7.  The frequency with which a  
feature with S/N = 5 appears in our spectra is $\simeq$0.02\%, which 
corresponds to just under a 4$\sigma$ detection for a Gaussian distribution.  
Though the frequency of features with fluxes comparable to our candidates may 
be slightly higher than indicated by Gaussian statistics, the test clearly 
illustrates that features of prominence similar to those of our candidates 
are extremely rare in clean regions of the spectra. 

\section{Verifying the Candidate Redshifts}

If the emission features discussed above are Lyman-$\alpha$ , then the source 
redshifts for our 6 candidates range between $z=8.7$ and $z=10.2$ and the 
observed fluxes span 2-5$\times10^{-17}$ erg cm$^{-2}$ s$^{-1}$.  Converting 
these observed fluxes into intrinsic luminosities is complicated by 
uncertainties in the fraction of emitted Lyman-$\alpha$ photons that reach 
the observer. We can compute a lower limit to the unlensed Lyman-$\alpha$ 
luminosity by converting the fluxes to luminosities and dividing by the 
magnification factor.  With these assumptions, the derived 
unlensed luminosities range from 1.6-50$\times$10$^{41}$ erg s$^{-1}$.  

We convert these Lyman-$\alpha$ luminosities to star formation rates
assuming case B recombination: $\rm L_{\alpha} = \frac{2}{3}h\nu_{Ly\alpha} 
(1-f_{esc}) Q(H) \dot{M_\star}$ where $\rm h\nu_{\alpha}$ is the energy of a 
Lyman-$\alpha$ photon, Q(H) is the Hydrogen ionizing photons flux, and 
$f_{esc}$ is the fraction of ionizing photons that escape from the galaxy 
(and hence are not emitted in recombination emission lines).  Assuming a  
Salpeter initial mass function, 1/20th solar metallicity, and a minimum 
and maximum stellar mass of 1 and 20 M$_\odot$, respectively, the hydrogen 
ionizing photon flux is 4$\times$10$^{53}$ s$^{-1}$ \citep{Schaerer03} for 
galaxy forming stars continuously at a rate of 1 M$_\odot$ yr$^{-1}$.
For $f_{esc} \ll 1$ (e.g. \citealt{Shapley06}), this results in a 
conversion factor of 
1 M$_\odot$ yr$^{-1}$ = 4.2$\times$10$^{42}$ erg s$^{-1}$.  With these 
assumptions, the star formation rates range from 
0.038 to 1.2 M$_\odot$ yr$^{-1}$.  If a Scalo IMF was adopted instead, the 
star formation rates would be a factor of three greater \citep{Loeb2005}.

We note that these star formation rates are uncertain for several 
reasons.  If dust in the galaxy absorbs any Lyman-$\alpha$ 
photons, then the intrinsic luminosities and corresponding star formation 
rates will be larger.  Likewise, if only a fraction of the Lyman-$\alpha$ 
line is transmitted through the IGM due to the presence of neutral 
hydrogen, then Lyman-$\alpha$ luminosity and star formation rates 
will again be underestimated.  On the other hand, if the stellar IMF at 
these early times is more top-heavy than a Salpeter or Scalo IMF, more 
ionizing photons are predicted per unit of star formation.  Similarly,
if the metallicity of these galaxies is less than what is assumed 
above, the efficiency of ionizing photon production (and hence 
Lyman-$\alpha$ luminosity) is increased \citep{Tumlinson00,Tumlinson01,
Schaerer03}.  Hence, in these cases, the star formation rates tabulated 
will be overestimates.

But how can we be sure, with a single emission line, that we have truly 
detecteda Lyman-$\alpha$ emitter?  Normally, for a narrow-band imaging survey 
\citep{Hu02,Hu04,MR04,Shimasaku06,Kashikawa06} or an unlensed 
spectroscopic survey \citep{Martin06}, contamination from foreground [O II]
and H$\alpha$ emitters can be significant. However, it is worth pointing out 
that such foreground contamination is likely to be much reduced for a 
critical-line survey designed to probe faint emitters since, for lensing 
clusters at $z\simeq $0.2-0.5, {\em all contaminating emitters will also be 
lensed and thus their respective most-likely location in the image plane will 
be spatially offset with respect to that for a source at $z\simeq $2}. The 
degree to which this benefits a lensed survey with respect to a blank field 
survey will depend on the relative magnifications at each redshift and the 
shape of the faint end of the respective line emitting luminosity
functions.

A redshift that the general community normally accepts is usually one in which
at least two, preferably more, features are reliably identified. Single line
identifications are quite naturally questioned as unreliable. As we probe to
$z\simeq 10$ with ground-based telescopes, however, we reach
the situation where multiple-line identifications are simply not practical.
With Lyman-$\alpha$ in the J-band no additional feature is visible in the 
full practical spectroscopic range from the UV limit to the beginning of the 
thermally-dominated background at 2$\mu$m.  The Infrared Spectrograph 
({\em IRS}) on Spitzer in principle offers the possiblity of detecting 
H$\alpha$ emission from $z\simeq 10$ objects. However, sensitivity 
calculations (Lacy et al., in prep.) suggest that the H$\alpha$/Lyman-$\alpha$ 
flux ratio would have to be significantly greater than that expected for 
case B recombination (perhaps due to Lyman-$\alpha$ suppression from 
resonant scattering and absorption) in order for this to be an effective 
verification.  

A second-best verification of Lyman-$\alpha$ that might be convincing would 
be the identification of an associated Lyman break in broadband imaging of 
the continuum distribution \citep{Stanway04} or the presence of an asymmetric 
line profile \citep{Ellis01,Hu02}. The Lyman break is due to intergalactic 
absorption of flux shortward of the Lyman-$\alpha$ line and causes all 
$z > 7$ sources to be undetected at optical wavelengths.  However, the lack 
of an optical broad-band detection need not necessarily rule out a lower 
redshift source, for example those with very strong emission line spectra 
and weak continuum emission.

Asymmetric line profiles are expected because of resonant scattering by 
neutral hydrogen within the host galaxy and in the IGM. This leads to 
absorption on the blue side of the line as the light redshifts into resonance, 
leaving a sharp cutoff on the blue side of the line and an extended red 
damping wing. Unfortunately, the detection of an asymmetric line profile 
requires very long integrations and high spectral resolution in order to 
detect the damping wing. 

A key factor in our consideration of how to verify the Lyman-$\alpha$ label 
for our 6 candidates is the importance of exquisite conditions to see such 
faint emitters. The required seeing and transparency occur only 40-50\% of 
the time even on an excellent site such as Mauna Kea. Conservatively, we 
estimated that to measure the Lyman-$\alpha$ profile of any one of our 6 
candidates with adequate signal/noise would take 8-10 hours of integration 
on Keck II. Allowing for weather, to follow-up all 6 candidates in this 
fashion would take 24 nights; clearly this is not a practical proposition!

One might conclude we have reached new territory where we may never 
quite know with certainty whether an object is at $z\simeq 10$. In practice,
however, there is a way forward although it is less definitive and more
statistical in nature. The approach we will adopt is to search 
spectroscopically in various passbands for lines associated with alternative, 
lower redshift explanations for the detected J-band emission line.  In ruling 
out a particular alternative identification, we must make some assumptions 
about the likely emission line spectrum of the interloper.  As such this is 
then a statistical process since we cannot cover all possible emission line 
ratios.  While this method will never confirm the Lyman-$\alpha$ 
interpretation with 100\% certainty, it is clearly an essential pre-requisite 
to spending 2 nights per candidate measuring the line profiles.

In the following subsections, we proceed to apply the tests discussed above to 
our candidates in order to constrain their redshifts.  In $\S$5.1, we examine 
the stacked line profiles of the six candidates to see if we can collectively 
assess the likelihood the bulk represent Lyman-$\alpha$ emission; in $\S$5.2, 
we test for the presence of a Lyman break in deep broadband photometry; 
finally in $\S$5.3, we discuss our follow-up spectroscopic program to test 
for low-redshift interlopers. 

\subsection{Stacked Line Profile of Candidates}

Asymmetric line profiles cannot be discerned in the individual emission 
features, because of the low spectral signal/noise.  As discussed, much 
longer integrations (8-10 hrs) would be needed to characterize the emission 
line profiles and this seems pointless until lower redshift interlopers have 
been rigorously tested.

Nonetheless, one can ask whether statistically as a combined set, the combined
line profile of all 6 candidates reveals an asymmetric profile suggestive of a 
Lyman-$\alpha$ origin.  The practicality of such an exercise is limited by a 
number of factors.  First the resolution of NIRSPEC is only moderate; 
$R\simeq $1500 c.f. $R\simeq $15,000 for the ESI spectrum obtained at $z$=5.7 
\citep{Ellis01}.  Second, stacking assumes that all 6 lines are 
Lyman-$\alpha$ with similar profiles and, most importantly, requires an 
accurate registration using the central wavelength within each low 
signal/noise profile. Finally, as the adjacent OH sky lines are distributed 
differently for each emitter,  wavelength regions that are clean in one 
candidate become averaged with noisy regions in another so the
contrast of the stacked line does not increase as $\sqrt{N}$. Even with 
weighted addition, we did not find the test to be very conclusive. 

\subsection{Searching for Lensed Pairs}

In several of the lensed emitters and drop-outs we have published in the
past \citep{Ellis01,Kneib04}, the location of a second or third
counter-image has been a particularly convincing demonstration of both
the lensed hypothesis and, via the mass model, the approximate redshift.
Indeed, the accurate recovery of the third image for the $z\simeq 6.8$
drop-out source in Abell 2218 \citep{Kneib04} was one of the major
arguments justifying its location beyond $z\simeq$6.

Generally speaking, image pairs are expected in strong lensing
configurations depending on the degree of alignment between the
background source and the cluster caustics, the physical size of
the source and, in this case, the ground-based seeing. The key
feature which led to the satisfactory recovery of image pairs in
our earlier studies was a broad-band detection with HST.
As an example, in the case of the $z=5.7$ pair discussed by
Ellis et al (2001), only a single emitter was recovered by LRIS.
The location of the second image required a detection by HST.
Indeed, pairs were not detected for any of the other lensed
sources in the subsequent LRIS survey \citep{Santos04}.

Some true pairs may simply not be resolved in our ground-based
spectra and, quite possibly, the second image lies outside our
NIRSPEC slit coverage. For the most well-constrained mass model, the
location of a counter-image can usually be predicted with an uncertainty
of 0\farcs5-1\farcs, i.e. to within 1-2 NIRSPEC slit widths. Although it
is practical to consider chasing the second images, this could
still be fairly time-consuming and we concluded that priority should be
given in eliminating low-redshift interlopers from our catalog of candidates.

\subsection{Broadband Photometry of Candidates}

The next test is to determine whether optical continuum emission is visible
for any of our candidates. Such a detection would clearly rule out a high z
interpretation although non-detection does not imply the opposite. Deep
ACS/WFPC-2 \citep{Kneib04,Smith02} images are 
available for all 9 clusters so we have examined
these images at the precise locations of each of our 6 emitters.

No definitive broadband optical detection is seen at the location of 
any of the candidates to R$\simeq $27.5 (3$\sigma$, Figure 8).  One 
of the candidates (Abell 2219 c2) is very marginally detected in the 
$z'_{F850LP}$-band; if this flux is associated with the J-band emission 
feature, then the high redshift interpretation would seem unlikely.  
Although 3 of the candidates (Abell 2219 c1, Abell 1689 c1, c2) 
are located within 2\farcs\~of brighter objects, in all cases, 
our registration sufficiently accurate to rule out the possibility that 
the J-band emission emanates from the bright objects. 

A detection in deep near-infrared imaging would be more interesting.
Coupled with the optical non-detections discussed above, this might
hint at the presence of a Lyman break, lending credence to a $z\simeq $10
interpretation. In fact, none of the candidates is detected at or above the 
5$\sigma$ level ($\simeq $27 AB mag) in deep broadband HST J$\rm_{110W}$, 
H$\rm_{160W}$ and Keck $J$-band imaging. Abell 2219 c2 is marginally 
detected at $\sim\,2\sigma$ in the H$\rm_{160W}$-band. 

In fact the lack of definitive near-infrared detections for our Lyman-
$\alpha$ candidates is not at all surprising if they are at high-z. If the 
equivalent width distribution of Lyman-$\alpha$ emitters at $z\simeq 10$ is 
comparable to that at $z\simeq $4-6, then we would need to reach 
$H_{160W}=27.7$ (1 magnitude below our sensitivity limit) to detect our 
candidates, if they are indeed Lyman-$\alpha$ emitters \citep{MR04,
Shimasaku06}.  

Thus, we can conclude that our deep imaging adds a further important
component in the case for the hypothesis that bulk of our emission line
sources lie at $z\simeq $10.

\subsection{Spectroscopically Testing Low Redshift Scenarios}

Aside from Lyman-$\alpha$, the most likely alternative identifications 
for the J-band emission features include H$\alpha$, [OIII], H$\beta$, and 
[OII], in which case the features would correspond to lensed sources at 
$z=0.5-2.5$.  Sources located at these redshifts would almost certainly have 
additional, associated emission lines bright enough to be seen in the optical 
and near-infrared spectral regions (see Table 4).  We have constrained each 
of these lower redshift hypotheses by undertaking further spectroscopy, 
placing stringent limits on the presence of lines.

This technique has value so long as the expected line spectrum of the
foreground interlopers is known. In practice selective extinction and 
excitation differences make the relevant line ratios somewhat uncertain. 
The best we can do is to approach the problem statistically, assuming 
a typical HII region spectrum (Figure 9) allowing some leeway in 
the line ratios.  

We generate the HII region spectrum by computing flux ratios of bright 
recombination lines (Lyman-$\alpha$, H$\alpha$, and H$\beta$) assuming 
Case B recombination and accounting for the possibility that a signficant 
fraction of Lyman-$\alpha$ photons are absorbed by dust thereby reducing 
its flux by up to a factor of three. With these assumptions, we 
adopt the following emission line ratios for hydrogen recombination lines: 
Lyman-$\alpha$ : H$\alpha$ : H$\beta$ = 7.4-22 : 2.8 : 1.0.  Flux ratios 
involving the forbidden oxygen lines ([OII] and [OIII]) can be determined 
from observations of low- and high-metallicity galaxies (Pilyugin et al. 
2000, 2001).  Here we adopt a ratio of [OIII]$\lambda5007$ : [OIII]$
\lambda4959$ : [OII] : H$\beta$ = 0.8-7 : 0.3-2 : 0.8-3 : 1.  Using 
these ratios and the conversion between Lyman-$\alpha$ luminosity and 
star formation rate (\S5), we generate a template spectrum (Figure 9);
the continuum flux spectrum is determined for a given star formation rate 
via a \cite{bc03} model with identical stellar properties to those assumed 
in \S5.

There are certainly a number of shortcomings in such an approach.  
Predicting the flux of Lyman-$\alpha$ is always difficult 
due to the resonant nature of the transition.  While we allow for a
factor of three suppression in the Lyman-$\alpha$ flux, it is possible 
that a larger fraction of Lyman-$\alpha$ photons could be absorbed.   
Furthermore, the flux ratios, especially those involving the 
forbidden oxygen lines, are dependant on the metallicity and the 
effective temperature of the ionizing stars.  However, as we will show 
below, our observational flux limits are often sufficiently tight to rule out 
low-redshift hypotheses for a wide variety of metal abundance and 
stellar effective temperature assumptions.  

Following the above logic, after completion of the blind scanning survey, we 
set out to secure additional optical and H-band spectroscopy for all 6 
candidates to apply this interloper-rejection method. Sadly, after two 
seasons of observing, weather and other observing vagaries mean that 
exhaustive consideration of all interloper possibilities is only available for 
3 of our candidates. For the other 3 candidates, only partial coverage
is in place.  

Table 4 summarizes the various hypotheses for the J--band emission and 
lists, for each case, which additional lines would be expected and at what 
wavelengths.  Optical spectroscopy is very efficient at constraining 
the likelihood that our J-band emission arises from either $H\alpha$ 
(z$\simeq $0.9) or  
[O III] ($z\simeq $1.5) since, in these cases, the LRIS data probes 
very deep in the associated wavelength regions where [O II] would be seen 
(at 7100 \AA\ and 9300 \AA\ respectively).  In some cases, the survey J-band 
NIRSPEC data can also be used to constrain the hypothesis that the
primary detection is $H\beta$ or [O III] where we would expect to locate one 
or both of the [OIII] $\lambda$4959,5007 pair and H$\beta$.  In practice, this 
is complicated by the fact  that lines can be obscured by the OH night sky
or seen at very low significance (e.g. 2$\sigma$).  We found that  such cases 
can be more effectively dispensed with using H-band spectroscopy to 
search for H$\alpha$ which should be easily observable if it lies 
in between the sky lines.  

\subsection{Best Candidates: Abell 68 c1 and Abell 2219 c1}

The most convincing elimination of foreground interloper status has been
achieved for the two candidates Abell 68 c1 and Abell 2219 c1. In our 
subsequent analysis we will assume these emitters are Lyman-$\alpha$ and 
so we describe these spectroscopic constraints in some detail. 

In addition to the null broad-band detections discussed earlier, deep optical 
LRIS and H-band NIRSPEC spectroscopy was secured in good conditions
for both candidates and no definitive features are seen in either case. 
We summarize the predicted fluxes and limiting sensitivities for each 
low-redshift interloper case in Table 5. 

In the case of Abell 68, if the J-band line was H$\alpha$, for our adopted
template H II region spectra, [OII] and H$\beta$ would have been recovered
with $>45\sigma$ in the LRIS spectrum. 

If the J-band feature was [OII], then Lyman-$\alpha$ would have been seen with
ease in the blue LRIS spectrum.  While the resonant nature of the transition 
makes it difficult to predict the Lyman-$\alpha$ flux , if even 0.1\% of the 
photons can escape, a $\simeq$5$\sigma$ detection would still have resulted. 
Since Lyman-$\alpha$ emission is often not seen in distant Lyman break 
galaxies \citep{Shapley03}, we
can secure additional constraints from the non-detection of H$\beta$ and the 
two [OIII] lines in the H-band spectrum.  Although some of the relevant
regions are affected by OH emission, [OIII] 5007 \AA\  would be strong enough 
to be detected at $>$5$\sigma$ under some cases.

If the line were [OIII]$\lambda$5007\AA~, then H$\beta$, 
[OIII]$\lambda$4959\AA~, and H$\alpha$ would all fall in the spectral 
region covered by the H-band spectroscopy.  H$\beta$ and H$\alpha$ would 
each be seen at $>$5 $\sigma$ for most of the range of emission line ratios. 

One seemingly unlikely hypothesis deserves attention. The candidate is 
quite close (1.8 arcsec) to a $z$=1.58 galaxy with H$\beta$ and [OIII] 
revealed in the J-band discovery spectrum (Richard et al. 2006, 
in preparation). If the candidate line were H$\beta$, the 
rest-frame velocity separation between the two sources would only be 
400 km s$^{-1}$.
The likelihood of an association thus seems high. However, it turns out the
combined J+H spectra of the two sources are very different. H$\alpha$ is seen 
from the foreground galaxy in the H-band but the same line can be rejected 
for the candidate with 15$\sigma$ confidence for a Case B Balmer line ratio.
Likewise the limiting sensitivity at the location of [OIII] emission line 
for the foreground
hypothesis is sufficiently deep such that we would have expected to see a line 
for nearly all reasonable line ratio assumptions.  In combination, therefore, 
the non-detection of H$\alpha$ and [OIII]$\lambda$ 5007 \AA~ strongly rule 
out any association of the candidate with the z=1.58 galaxy.

Similarly, in the case of Abell 2219 c1 if the J-band line were H$\alpha$, 
[OII], H$\beta$ and [O III] 4959\AA~ would have been seen in the LRIS
spectra at 23-85, 8.5, and 6.4-43 $\sigma$ respectively. 

If the line was H$\beta$,  H$\alpha$ would have been detected at 22$\sigma$ 
in the H-band spectrum.and [OIII] $\lambda$4959\AA~ and $\lambda$5007\AA~ 
at 3.5-24 and 21-190 $\sigma$ respectively in the J-band.

If the line were [OII], then, as before strong Lyman-$\alpha$ would be expected
in the blue LRIS spectrum, even if only a small fraction of photons escaped.
And [OIII]$\lambda$4959\AA~ and $\lambda$5007\AA~ would have been
seen at up to 4.5-8.0 and 4.8-11$\sigma$ respectively in the H-band spectrum.

Finally, if the line were [OIII] $\lambda$5007\AA~, then H$\beta$ would
be detected at 1.7$-$15$\sigma$, depending on the oxygen abundance.
and H$\alpha$ would been seen in a clean area of the H-band with 
4.8$-$43$\sigma$.  

All cases described above assume an HII region template spectrum.
Assuming the emission lines are instead generated by a harder ionization 
source such as an AGN, the CIV $\lambda\lambda$ 1548, 1550
line may be prominent in the rest-frame UV spectrum.  If Lyman-$\alpha$ 
is suppressed via resonant scattering and absorption by dust, narrow-line 
AGN can exhibit line ratios of 1$\lsim$ CIV/Ly$\alpha$ $\lsim$ 1.4 
\citep{Dey95,Debreuck01,Dawson03}.  If these sources are indeed 
Lyman-$\alpha$ emitters at $z\simeq 10$, then the CIV line may also 
be observable in the H-band.  For both Abell 68 c1 and Abell 2219 c1, 
CIV would lie in clean regions (1.60 $\mu$m and 1.55 $\mu$m,
respectively); however, neither emission line is seen, suggesting that if the 
candidates are at $z\simeq 10$, they have CIV/Ly$\alpha <$ 1.

\subsection{Other Candidates}

Comprehensive optical and H-band spectroscopy is also available 
for Abell 2219 c2 and in fact the constraints we discuss above for 
Abell 2219 c1 are just as convincing here making it a very promising 
source. However, we prefer not to elevate it into the status of the two 
most promising sources because of the potential broad-band ACS 
detection (\S5.2).  In this sense it represents a source of status intermediate
between our top two candidates and those discussed below for
which, due to weather and other problems, our spectroscopy is
currently incomplete.

For A1689c1, c2 and c3, we have no optical or H-band spectroscopy
and so major interloper hypotheses such as $H\alpha$ and [O II]
cannot yet be excluded.  

In summary,  we can conclude that at least two (and quite possibly
three) of our six candidates cannot be accounted for as lines
from lower redshift emitters. Moreover, there is no reason to denigrate
the status of the other 3 candidates either since, so far, every 
additional spectrum we have taken has continued to support the 
Lyman-$\alpha$ hypothesis. Our low contamination ratio is not that surprising
when one considers all of the considered interloper hypotheses would
still represent lensed systems and thus, if highly magnified, not be
seen at the location of the $z\simeq $10 critical line.

On the other hand, it is clear that our tests make necessary
assumptions about the line ratios expected from $0.85<z<$2.26
galaxies. It is conceivable that these assumptions are
invalid and that the sources have anomalous line ratios (although it would 
seem surprising that this is the case in several systems). Given the 
wavelength ranges accessible to ground-based spectrographs, we 
believe we have undertaken all possible tests. 

\section{Implications}

Noting that at least two of our six candidates may indeed lie at $z\simeq $10,
we now address the goal of the survey: what is the volume density of high z
emitters? However, noting their identification could still be considered 
tentative, we will also explore the implications if {\em none} of our 
candidates is real. After making simple assumptions about the effect of 
dust and resonant scattering on escaping Lyman-$\alpha$ photons, we will 
use the derived abundance to constrain the star formation rate density.  
We then attempt to understand whether the derived abundance represents a
significant contribution to the ionizing photon budget at $z\simeq 10$.

\subsection{Volume Density of z$>$8 Lyman-$\alpha$ Emitters}

In a blank-field survey, the surface density of sources brighter than a given 
limiting line flux can be computed by dividing the number of detected sources 
by the area observed.  The calculation is more complex in this case because 
of the varying amplification of the source flux over the survey volume and 
the distortion, by lensing, of the survey area. 

The differential comoving volume element at ($\sky$,$z$) along one of 
the NIRSPEC slit positions is

\begin{equation}
\dd V_\mathrm{c}(\sky,z) = \frac{1}{\mg}
\left[ \frac{\dd l_\mathrm{c}(z)}{\dd z} \dd z \right]
\times \left[ D_\mathrm{c}^2(z) \dd\sky \right].
\end{equation}
where $\dd l_\mathrm{c}$ is the comoving
length of the volume element along the line of sight. We also 
compute the uncertainty associated with each volume element 
due to the error in the magnification ($\S$3.4).  

The total survey volume over which sources with Lyman-$\alpha$ line 
luminosity greater than $L$ is the integral over all volume 
elements with limiting Lyman-$\alpha$ luminosity below L,

\begin{equation}
V_\mathrm{c}(L) =
\int_\sky \int_z \dd V_\mathrm{c}(\sky,z)~H[ L-\llim ],
\end{equation}
where H[ L-$\llim$ ] is a step function equal to 1 when
L$\ge\llim$ and equal to 0 when L$<\llim$.

The uncertainty in the magnification introduces error into the 
derived comoving volume.  We quantify this error by computing the 
distribution of comoving volume for all acceptable cluster mass models 
(see $\S$ 3.4).  The range of comoving volume that spans $\pm$34\% of 
the total samples is taken to be the uncertainty in the volume.

The survey volume and associated errors are plotted as a function 
of Lyman-$\alpha$ luminosity in Figure 10.  The comoving volume decreases from 
$\simeq$30 Mpc$^3$ at L=3$\times$10$^{42}$ $\ergs$ to $\simeq$3 Mpc$^3$ 
at L=3$\times$10$^{41}$ $\ergs$.  The 'one-sigma' error bars in the volume are 
small enough that they do not extend outside the datapoints.
In Figure 11, we plot the number density of $z\simeq $9 Lyman-$\alpha$
emitters as a function of Lyman-$\alpha$ luminosity assuming that (i) 
none of our candidates is at  $z\simeq $8-10 (left) 
and (ii) two of our candidates are Lyman-$\alpha$ emitters (middle) and 
(iii) all six of our candidates are real (right).  

In the case where two of the candidates are real, the Poisson 
errors are 100\%.  This arises because galaxies need not only be 
brighter than the survey limiting luminosity but must also lie within 
the area covered by the survey.  In lensing experiments, in order 
for inclusion associated with some limiting luminosity, the 5$\sigma$ 
limiting luminosity at the location of the object must be lower than 
the limiting luminosity of the survey.  The net result is that objects 
with fluxes close to the sensitivity limits only are included in 
``surveys'' with a small range of limiting luminosities.  If two candidates 
are real, as considered here, the luminosity ranges over which the
objects would be detected do not overlap; hence no bin contains 
more than one object.

To place the inferred abundances in context, we compare
our results with other surveys that have been conducted,
particularly at z$\simeq$6 where the luminosity function of 
Lyman-$\alpha$ emitters is more firmly established.  

The first observational constraint on the abundance of Lyman-$\alpha$ emitters 
at z$>$8 was presented by \citep{Willis05}.  These authors conducted 
a narrow-band imaging survey in the J-band toward the Hubble Deep
Field South and found no $z=8.8$ sources with Lyman-$\alpha$ 
luminosity greater than 10$^{42.5}$ erg s$^{-1}$ over a volume of 
approximately 990 Mpc$^{-3}$. More recently, \cite{Cuby06} reported 
a null detection of $z=8.8$ Lyman-$\alpha$ emitters brighter than 
10$^{43}$ erg s$^{-1}$ over a comoving volume of $\simeq 4200 Mpc^3$.
Given the large error bars, the upper limits from \cite{Willis05} and 
\cite{Cuby06} are consistent 
with the case in which only two of the candidates are real.  If 
all six of the candidates are Lyman-$\alpha$ emitters, it becomes slightly 
more difficult to explain the lack of Lyman-$\alpha$ emitters in the two 
narrowband surveys without resorting to luminosity dependent 
evolution.  Hence, in this case, the observations could suggest that the 
decline in luminous Lyman-$\alpha$ emitters relative to fainter sources 
observed at $z\simeq 6-7$ \citep{Shimasaku06,Kashikawa06,Iye} 
continues to $z\simeq 10$.

A second question is whether our density is higher or lower
than the Lyman-$\alpha$ luminosity function which is now fairly 
well-established at $z=5.7$, at least at the bright end 
\citep{Santos04,MR04,Shimasaku06}. It is worth remembering that
despite the fact that over 100 Lyman-$\alpha$ emitters are now catalogued at
this redshift, the degree of spectroscopic confirmation is still quite
limited. The extant data are well-fit by a Schechter function 
with $\phi^\star$=1.6$\times$10$^{-4}$ Mpc$^{-3}$ and 
L$^\star$=1.6$\times$10$^{43}$ erg s$^{-1}$ where, for the purposes
of discussion, we have adopted the steepest likely faint-end slope of
$\alpha=-2$ so as to provide the most optimisitc no-evolution scenario.  
Assuming, naively, no evolution between z=5.7 and z=10, it is 
unlikely that we would find a source brighter than 10$^{40}$ erg s$^{-1}$.
At face value, therefore, our detection of at least two promising cases 
would suggest an upwards evolution in the abundance of faint emitters
with redshift.

The likely evolution of the Lyman-$\alpha$ luminosity function between  
z=5.7 and z$\simeq$10 is a complicated function of the star formation 
efficiency, stellar initial mass function, relative geometry and kinematics 
of HII and HI gas, and  dust content of the host galaxy, as well as the 
ionization state of the IGM.  If, for example, the dust content of galaxies 
decreases toward higher redshifts, the percentage of Lyman-$\alpha$ photons 
that escape the host galaxy may increase. On the other hand, as the IGM 
becomes increasingly neutral, observed Lyman-$\alpha$ counts are expected to 
decline \citep{MR04,Haiman05}, although the presence of large HII regions 
around clustered star-forming sources may slightly offset the magnitude of 
this decline \citep{Furlanetto04}. Clearly, it is very difficult to predict 
exactly {\it how} the Lyman-$\alpha$ luminosity function will evolve over 
this $\simeq$500 Myr time interval.  

Bearing these uncertainties in mind, our detection of two promising
candidates clearly implies a steep faint-end slope and hence a large 
abundance of intrinisically faint sources. If more of our candidates 
are at $z\simeq 10$, it would almost certainly imply that low-luminosity 
star-forming systems are {\it more} abundant at $z=9$ than at $z=5.7$.  
This may be a result of evolution in galaxy properties (as discussed),
or the effect of a predominantly neutral IGM which increases star formation 
efficiency in low-mass halos due to the decrease in the Jeans mass  
\citep{BL01}. Determining which of these factors is primarily 
responsible for the possible upturn in counts is difficult with the 
available data.  
  
If none of our candidates is real, a number of explanations 
are possible.  First the faint-end slope may not be as steep 
as required for us to detect sources in our observed volume.  
There are a number of feedback processes (galactic winds, 
supernova explosiona, etc.) that could significantly reduce star 
formation efficiency in low-mass halos, thereby resulting in a 
relatively shallow faint-end slope.  Alternatively, if the 
IGM is significantly neutral, it is possible
that faint-end slope is very steep, but the neutral IGM is blocks our 
view of these systems.  We defer a detailed discussion of the
implications of our survey in terms of what is now known at $z\simeq $5-6
to a later paper (Stark, Loeb \& Ellis, 2007).

\subsection{Contribution of Low Luminosity Galaxies to Reionization}

Several methods have been adopted for considering the contribution of 
star-forming sources to cosmic reionization.  One approach is to compare
the inferred ionizing photon production rate from star-forming galaxies 
of all luminosities to that required to reionize neutral hydrogen in the 
IGM \citep{Madau99,Sti04a}.  Since the abundance of Lyman-$\alpha$ 
emitters at $z\simeq 10$ is very uncertain, comparing the total ionizing 
photon output is highly dependent upon assumptions about the luminosity 
function. In this initial study, we ask a more basic question: do low 
luminosity Lyman-$\alpha$ emitters provide a significant contribution 
to reionization? Following this approach, we therefore only tabulate 
the ionizing photon production arising from low luminosity (i.e. 
10$^{41}$ - 10$^{42}$ erg s$^{-1}$) sources that our survey is designed 
to probe.   
 
The contribution of low luminosity star-forming sources to 
reionization can be parameterized in simple terms by comparing their  
ionizing photon production rate to that required for reionization:
\begin{eqnarray}
\rm n_{gal} =  2~\Big(\frac{B}{10}\Big) \Big(\frac{f_{esc}}{0.05}\Big)^{-1} \Big(\frac{\dot{M_\star}}{0.1~M_\odot~yr^{-1}}\Big)^{-1} \nonumber \\
\rm \Big(\frac{\Delta t}{575~Myr}\Big)^{-1} ~Mpc^{-3}
\end{eqnarray}
assuming the space density of sources is constant during the reionization
epoch.  Here, B is the number of ionizing photons required to keep a
single hydrogen atom ionized, n$\rm_H$ is the comoving number density of 
hydrogen at the redshift of interest, $\Delta$t is the period over which 
reionization occurs, f$_{esc}$ is the escape fraction of ionizing photons, 
$\rm\dot{M_\star}$ is the time-averaged star formation rate of galaxies in 
units of M$_\odot$ yr$^{-1}$. 

In both cases, we fix the comoving number density of hydrogen at 
n$\rm_H$=1.9 $\times$ 10$^{-7}$ cm$^{-3}$, appropriate
for the comoving hydrogen number density adopting the best-fit WMAP
cosmological parameters \citep{Sp06}.  We have also assumed 
an IMF with Salpeter slope with stellar masses ranging from 1 M$_\odot$ 
to 100 M$_\odot$ and metallicity of of Z=0.001 (1/20th solar) 
\citep{Schaerer03}. If the IMF is weighted more toward massive stars or is 
substantially more metal-poor, the ionizing photon output for a given star 
formation rate will be greater, and less galaxies will be required for 
reionization.  

The escape fraction of ionizing photons is not well-constrained 
observationally; we allow this quantity to vary between 0.02 and 0.5 
\citep{Steidel01,FSoto03,Shapley06}.  We assume the escape 
fraction is constant with luminosity; however, models of high-redshift 
galaxies suggest that it should increase toward the low-luminosity 
end of the luminosity function \citep{Ricotti00}.  If this is the 
case, then contribution of low-mass galaxies to reionization will 
be even greater than what is shown below.  

The recombination rate increases with the local hydrogen number
density, hence the number of ionizations required per atom for
reionization depends strongly on the clumpiness of the IGM
(C=$\frac{<n_e^2>}{<n_e>^2}$).  Simulations generally predict clumpiness
factors of C$\sim$30 \citep{Gnedin2000}
for z$\sim$6-7.  Following the interpolation formulae in \citep{Sti04a}, this
results in $\simeq$5-10 ionizations per hydrogen atom depending on the 
timescale of reionization.  

Substituting the range of parameter values into equation 5, we find that 
for galaxies with star formation rates of 0.1 M$_\odot$ yr$^{-1}$ 
to dominate the reionization process, they should have comoving 
number densities in excess of 0.1 - 4 Mpc$^{-3}$ (Figure 12).  
If two or more of the candidates are at z$\simeq $10, then it would 
appear that low luminosity galaxies contribute significantly 
to the reionization process.

While there is currently not enough data at $z\simeq 10$ to estimate 
the integrated ionizing photon output from Lyman-$\alpha$ emitters, 
we can attempt to constrain this quantity by assuming that the 
luminosity function evolves only in normalization (hence maintaining
the same shape) between $z=6.5$ and $z=10$.  In the most optimistic 
case where six candidates are at high-redshift, we take the characteristic 
luminosity, L$^\star$ derived for the $z=6.5$ Lyman-$\alpha$ emitters 
by \cite{Kashikawa06} with a diverging 
faint-end slope of $\alpha=-2$.  Integrating the luminosity function down to 
the survey limit ($\simeq $0.01 M$_\odot$ yr$^{-1}$), we find that the total 
ionizing photon production rate would be sufficient for reionization for 
nearly the full range of assumptions considered in Figure 12.

\section{Summary}

Low luminosity star-forming galaxies are often predicted 
to produce a large supply of ionizing photons during the 
reionization era \citep{BL01}.  However, feedback processes
from supernova explosions or galaxy winds may significantly 
decrease star formation efficiency in the low-mass dark matter
halos that are thought to house the majority of low-luminosity 
star forming sources.  Tentative observational 
evidence hints at a very large abundance of low luminosity 
sources \citep{yw04,Bou06} suggesting that these feedback 
processes may not significantly affect star formation in low-mass halos. 
Until the construction of JWST and 20-40 meter class 
ground-based telescopes, gravitational lensing surveys offer one of the 
most efficient 
means of spectroscopically verifying the presence of an abundant population 
of low-luminosity ($\rm \lsim 10^{43} erg s^{-1}$) sources.  With this as 
our goal, we have conducted a spectroscopic survey for gravitationally 
lensed Lyman-$\alpha$ emitters over the redshift interval 8.5$<z<$10.4.  
We summarize our results below.

1. The critical lines of 9 well-understood lensing clusters were 
scanned spectroscopically in the J-band for lensed Lyman-$\alpha$ emission 
from galaxies at $z$=8.5 to 10.4. The new observations are 
sensitive to star-forming sources with intrinsic (unlensed) Lyman-$\alpha$ 
luminosities of 10$^{41.5}$ erg s$^{-1}$ over a significant fraction of 
the survey area.  This limit is over an order of magnitude fainter than
the corresponding sensitivity limits of conventional surveys for 
Lyman-$\alpha$ emitters at $z>6$.

2. Six promising candidate Lyman-$\alpha$ emitters were identified whose
reality we justify noting the possibility of instrumental and
detector artefacts.  The emission features have fluxes that range 
from 2-5$\times10^{-17}$ erg cm$^{-2}$ s$^{-1}$ and significances 
between 5-8$\sigma$.  The Lyman-$\alpha$ luminosities of the sources range 
from 10$^{41}$ to 5$\times$10$^{42}$ erg s$^{-1}$.  

3. We attempt to verify that our lines arise from $z\simeq $10 \lya
via several tests. No convincing broadband optical counterparts 
were seen in deep HST images (R$\simeq $27) except in one
marginal case. For 3 out of the 6 cases, we conduct exhaustive
additional spectroscopy to explore the likelihood that low-redshift 
interloper lines may explain the J-band emission. In no case
do we find such an example, and, overall we conclude that at least
2 of our 6 candidates probably lie at $z\simeq $10.  We briefly
discuss the considerable challenges of making further progress
in confirming any or all of our candidates given the limited
rest-wavelength range available with large ground-based
telescopes. 
 
4. Assuming two or more of our Lyman-$\alpha$ emitter candidates are real, 
then the cumulative abundance of low luminosity galaxies (defined as 
those with L$>$10$^{41.5}$ erg s$^{-1}$) is at least 0.3 Mpc$^{-3}$.
Such a large abundance of low luminosity Lyman-$\alpha$ emitters supports
the contention of a steep faint end slope for the star-forming
luminosity function at z$\simeq$10.

5. Assuming a single population of galaxies dominates the 
reionization process the number density of galaxies forming 
stars at a rate of 0.1 M$_\odot$ yr$^{-1}$ must exceed
0.1-4 Mpc$^{-3}$ to reionize neutral hydrogen.  Our inferred 
abundance is consistent with being within this range indicating 
that intrinsically faint Lyman-$\alpha$ emitters may well produce a 
substantial supply of ionizing photons during the reionization era. 

6. However, if none of our candidates are at high-redshift, then there are
 several likely explanations. If  the IGM is significantly neutral, 
 low luminosity star forming systems may be very abundant but 
 may not be observed due to resonant absorption of Lyman-$\alpha$ photons 
 by hydrogen in the IGM.  Alternatively, it is possible that the 
 faint-end slope is not as steep as predicted by \cite{yw04}; 
hence larger volumes would be needed to detect 
a representative sample.

Notwithstanding the uncertainties and speculation, our survey
demonstrates the practicality, over the next few years, of
providing a valuable glimpse at the nature of the $z\simeq $10
Universe ahead of the commissioning of future large facilities
such as the Thirty Meter Telescope and the James Webb
Space Telescope.

\acknowledgments

We thank our anonymous referee and Avi Loeb for very helpful comments.  
We are indebted to George Becker for providing 
his spectroscopy reduction software and Tom Broadhurst for providing 
mass models for several of our survey clusters.  We thank Mark Sullivan 
for observing and reducing data on the candidate Abell 68 c1 with LRIS.  
G.P.S. acknowledges support from a Royal Society University 
Research Fellowship.  Faint object near-infrared spectroscopy 
at Keck is possible thanks to the dedicated efforts of 
instrumentalists and the Keck support staff; we thank Ian McLean, 
Jim Lyke, and Grant Hill for making this project possible.

The authors recognize and acknowledge the very significant cultural
role and reverance that the summit of Mauna Kea has always had
within the indigenous Hawaiian community.  We are most fortunate
to have the opportunity to conduct observations from this mountain.

\bibliography{journals_apj,mybib}

\begin{thebibliography}{}

\bibitem[\protect\citeauthoryear{{Barkana} \& {Loeb}}{{Barkana} \&
  {Loeb}}{2001}]{BL01}
{Barkana}, R.,  \& {Loeb}, A. 2001, \physrep, 349, 125

\bibitem[\protect\citeauthoryear{{Beckwith} et~al.}{{Beckwith}
  et~al.}{2006}]{Beckwith06}
{Beckwith}, S.~V.~W., et~al. 2006, \aj, 132, 1729

\bibitem[\protect\citeauthoryear{{B{\'e}zecourt} et~al.}{{B{\'e}zecourt}
  et~al.}{1999}]{Bez99}
{B{\'e}zecourt}, J., {Kneib}, J.~P., {Soucail}, G.,  \& {Ebbels}, T.~M.~D.
  1999, \aap, 347, 21

\bibitem[\protect\citeauthoryear{{Borys} et~al.}{{Borys}
  et~al.}{2004}]{Borys04}
{Borys}, C., et~al. 2004, \mnras, 352, 759

\bibitem[\protect\citeauthoryear{{Bouwens} \& {Illingworth}}{{Bouwens} \&
  {Illingworth}}{2006}]{Bou06b}
{Bouwens}, R.~J.,  \& {Illingworth}, G.~D. 2006, \nat, 443, 189

\bibitem[\protect\citeauthoryear{{Bouwens} et~al.}{{Bouwens}
  et~al.}{2006}]{Bou06}
{Bouwens}, R.~J., {Illingworth}, G.~D., {Blakeslee}, J.,  \& {Franx}, M. 2006,
  \apj,submitted

\bibitem[\protect\citeauthoryear{{Bouwens} et~al.}{{Bouwens}
  et~al.}{2005}]{Bou05}
{Bouwens}, R.~J., {Illingworth}, G.~D., {Thompson}, R.~I.,  \& {Franx}, M.
  2005, \apjl, 624, L5

\bibitem[\protect\citeauthoryear{{Bremer} et~al.}{{Bremer}
  et~al.}{2004}]{Bremer04}
{Bremer}, M.~N., {Jensen}, J.~B., {Lehnert}, M.~D., {Schreiber}, N.~M.~F.,  \&
  {Douglas}, L. 2004, \apjl, 615, L1

\bibitem[\protect\citeauthoryear{{Broadhurst} et~al.}{{Broadhurst}
  et~al.}{2005}]{Broadhurst05a}
{Broadhurst}, T., et~al. 2005, \apj, 621, 53

\bibitem[\protect\citeauthoryear{{Bruzual} \& {Charlot}}{{Bruzual} \&
  {Charlot}}{2003}]{bc03}
{Bruzual}, G.,  \& {Charlot}, S. 2003, \mnras, 344, 1000

\bibitem[\protect\citeauthoryear{{Cuby} et~al.}{{Cuby} et~al.}{2006}]{Cuby06}
{Cuby}, J.~., {Hibon}, P., {Lidman}, C., {Le Fevre}, O., {Gilmozzi}, R.,
  {Moorwood}, A.,  \& {van der Werf}, P. 2006, Accepted for publication in
  A\&A, astro-ph/0611272

\bibitem[\protect\citeauthoryear{{Dawson} et~al.}{{Dawson}
  et~al.}{2003}]{Dawson03}
{Dawson}, S., {McCrady}, N., {Stern}, D., {Eckart}, M.~E., {Spinrad}, H.,
  {Liu}, M.~C.,  \& {Graham}, J.~R. 2003, \aj, 125, 1236

\bibitem[\protect\citeauthoryear{{De Breuck} et~al.}{{De Breuck}
  et~al.}{2001}]{Debreuck01}
{De Breuck}, C., et~al. 2001, \aj, 121, 1241

\bibitem[\protect\citeauthoryear{{Dey}, {Spinrad}, \& {Dickinson}}{{Dey}
  et~al.}{1995}]{Dey95}
{Dey}, A., {Spinrad}, H.,  \& {Dickinson}, M. 1995, \apj, 440, 515

\bibitem[\protect\citeauthoryear{{Ebbels} et~al.}{{Ebbels}
  et~al.}{1998}]{Ebbels98}
{Ebbels}, T., {Ellis}, R., {Kneib}, J.-P., {Leborgne}, J.-F., {Pello}, R.,
  {Smail}, I.,  \& {Sanahuja}, B. 1998, \mnras, 295, 75

\bibitem[\protect\citeauthoryear{{Ebeling}, {Edge}, \& {Henry}}{{Ebeling}
  et~al.}{2001}]{Ebeling01}
{Ebeling}, H., {Edge}, A.~C.,  \& {Henry}, J.~P. 2001, \apj, 553, 668

\bibitem[\protect\citeauthoryear{{Egami} et~al.}{{Egami} et~al.}{2005}]{Eg04}
{Egami}, E., et~al. 2005, \apjl, 618, L5

\bibitem[\protect\citeauthoryear{{Ellis} et~al.}{{Ellis}
  et~al.}{2001}]{Ellis01}
{Ellis}, R., {Santos}, M.~R., {Kneib}, J.-P.,  \& {Kuijken}, K. 2001, \apjl,
  560, L119

\bibitem[\protect\citeauthoryear{{Eyles} et~al.}{{Eyles} et~al.}{2006}]{Ey06}
{Eyles}, L.~P., {Bunker}, A.~J., {Ellis}, R.~S., {Lacy}, M., {Stanway}, E.~R.,
  {Stark}, D.~P.,  \& {Chiu}, K. 2006, \mnras, 1348

\bibitem[\protect\citeauthoryear{{Eyles} et~al.}{{Eyles} et~al.}{2005}]{Ey05}
{Eyles}, L.~P., {Bunker}, A.~J., {Stanway}, E.~R., {Lacy}, M., {Ellis}, R.~S.,
  \& {Doherty}, M. 2005, \mnras, 364, 443

\bibitem[\protect\citeauthoryear{{Fan} et~al.}{{Fan} et~al.}{2006}]{Fan06}
{Fan}, X., et~al. 2006, \aj, 132, 117

\bibitem[\protect\citeauthoryear{{Fern{\'a}ndez-Soto}, {Lanzetta}, \&
  {Chen}}{{Fern{\'a}ndez-Soto} et~al.}{2003}]{FSoto03}
{Fern{\'a}ndez-Soto}, A., {Lanzetta}, K.~M.,  \& {Chen}, H.-W. 2003, \mnras,
  342, 1215

\bibitem[\protect\citeauthoryear{{Furlanetto}, {Zaldarriaga}, \&
  {Hernquist}}{{Furlanetto} et~al.}{2004}]{Furlanetto04}
{Furlanetto}, S.~R., {Zaldarriaga}, M.,  \& {Hernquist}, L. 2004, \apj, 613, 1

\bibitem[\protect\citeauthoryear{{Gnedin}}{{Gnedin}}{2000}]{Gnedin2000}
{Gnedin}, N.~Y. 2000, \apj, 542, 535

\bibitem[\protect\citeauthoryear{{Haiman} \& {Cen}}{{Haiman} \&
  {Cen}}{2005}]{Haiman05}
{Haiman}, Z.,  \& {Cen}, R. 2005, \apj, 623, 627

\bibitem[\protect\citeauthoryear{{Henry} et~al.}{{Henry}
  et~al.}{2007}]{Henry07}
{Henry}, A.~L., {Malkan}, M.~A., {Colbert}, J.~W., {Siana}, B., {Teplitz},
  H.~I., {McCarthy}, P.,  \& {Yan}, L. 2007, Accepted to ApJ Letters,
  astro-ph/0701045

\bibitem[\protect\citeauthoryear{{Horton} et~al.}{{Horton}
  et~al.}{2004}]{Horton04}
{Horton}, A., {Parry}, I., {Bland-Hawthorn}, J., {Cianci}, S., {King}, D.,
  {McMahon}, R.,  \& {Medlen}, S. 2004, in Ground-based Instrumentation for
  Astronomy. Edited by Alan F. M. Moorwood and Iye Masanori. Proceedings of the
  SPIE, Volume 5492, pp. 1022-1032 (2004)., ed. A.~F.~M. {Moorwood} \&
  M.~{Iye}, 1022

\bibitem[\protect\citeauthoryear{{Hu} et~al.}{{Hu} et~al.}{2005}]{Hu05}
{Hu}, E.~M., {Cowie}, L.~L., {Capak}, P.,  \& {Kakazu}, Y. 2005, in IAU Colloq.
  199: Probing Galaxies through Quasar Absorption Lines, ed. P.~{Williams},
  C.-G. {Shu}, \& B.~{Menard}, 363

\bibitem[\protect\citeauthoryear{{Hu} et~al.}{{Hu} et~al.}{2004}]{Hu04}
{Hu}, E.~M., {Cowie}, L.~L., {Capak}, P., {McMahon}, R.~G., {Hayashino}, T.,
  \& {Komiyama}, Y. 2004, \aj, 127, 563

\bibitem[\protect\citeauthoryear{{Hu}, {Cowie}, \& {McMahon}}{{Hu}
  et~al.}{1998}]{Hu98}
{Hu}, E.~M., {Cowie}, L.~L.,  \& {McMahon}, R.~G. 1998, \apjl, 502, L99

\bibitem[\protect\citeauthoryear{{Hu} et~al.}{{Hu} et~al.}{2002}]{Hu02}
{Hu}, E.~M., {Cowie}, L.~L., {McMahon}, R.~G., {Capak}, P., {Iwamuro}, F.,
  {Kneib}, J.-P., {Maihara}, T.,  \& {Motohara}, K. 2002, \apjl, 568, L75

\bibitem[\protect\citeauthoryear{{Iye} et~al.}{{Iye} et~al.}{2006}]{Iye}
{Iye}, M., et~al. 2006, \nat, 443, 186

\bibitem[\protect\citeauthoryear{{Kashikawa} et~al.}{{Kashikawa}
  et~al.}{2006}]{Kashikawa06}
{Kashikawa}, N., et~al. 2006, \apj, 648, 7

\bibitem[\protect\citeauthoryear{{Kelson}}{{Kelson}}{2003}]{Kelson03}
{Kelson}, D.~D. 2003, \pasp, 115, 688

\bibitem[\protect\citeauthoryear{{Kneib} et~al.}{{Kneib}
  et~al.}{2004}]{Kneib04}
{Kneib}, J.-P., {Ellis}, R.~S., {Santos}, M.~R.,  \& {Richard}, J. 2004, \apj,
  607, 697

\bibitem[\protect\citeauthoryear{{Kneib} et~al.}{{Kneib}
  et~al.}{1996}]{Kneib96}
{Kneib}, J.-P., {Ellis}, R.~S., {Smail}, I., {Couch}, W.~J.,  \& {Sharples},
  R.~M. 1996, \apj, 471, 643

\bibitem[\protect\citeauthoryear{{Kneib} et~al.}{{Kneib}
  et~al.}{2003}]{Kneib03}
{Kneib}, J.-P., et~al. 2003, \apj, 598, 804

\bibitem[\protect\citeauthoryear{{Labb{\'e}} et~al.}{{Labb{\'e}}
  et~al.}{2006}]{Labbe06}
{Labb{\'e}}, I., {Bouwens}, R., {Illingworth}, G.~D.,  \& {Franx}, M. 2006,
  \apjl, 649, L67

\bibitem[\protect\citeauthoryear{{Loeb}, {Barkana}, \& {Hernquist}}{{Loeb}
  et~al.}{2005}]{Loeb2005}
{Loeb}, A., {Barkana}, R.,  \& {Hernquist}, L. 2005, \apj, 620, 553

\bibitem[\protect\citeauthoryear{{Lord}}{{Lord}}{1992}]{Lord92}
{Lord}, S. 1992, NASA Tech. Rep. 103957

\bibitem[\protect\citeauthoryear{{Madau}, {Haardt}, \& {Rees}}{{Madau}
  et~al.}{1999}]{Madau99}
{Madau}, P., {Haardt}, F.,  \& {Rees}, M.~J. 1999, \apj, 514, 648

\bibitem[\protect\citeauthoryear{{Malhotra} \& {Rhoads}}{{Malhotra} \&
  {Rhoads}}{2004}]{MR04}
{Malhotra}, S.,  \& {Rhoads}, J.~E. 2004, \apjl, 617, L5

\bibitem[\protect\citeauthoryear{{Martin} et~al.}{{Martin}
  et~al.}{2006}]{Martin06}
{Martin}, C.~L., {Sawicki}, M., {Dressler}, A.,  \& {McCarthy}, P.~J. 2006, New
  Astronomy Review, 50, 53

\bibitem[\protect\citeauthoryear{{McLean} et~al.}{{McLean}
  et~al.}{1998}]{Mclean98}
{McLean}, I.~S., et~al. 1998, in Proc. SPIE Vol. 3354, p. 566-578, Infrared
  Astronomical Instrumentation, Albert M. Fowler; Ed., ed. A.~M. {Fowler}, 566

\bibitem[\protect\citeauthoryear{{Mobasher} et~al.}{{Mobasher}
  et~al.}{2005}]{M05}
{Mobasher}, B., et~al. 2005, \apj, 635, 832

\bibitem[\protect\citeauthoryear{{Pell{\'o}} et~al.}{{Pell{\'o}}
  et~al.}{2004}]{Pello04}
{Pell{\'o}}, R., {Schaerer}, D., {Richard}, J., {Le Borgne}, J.-F.,  \&
  {Kneib}, J.-P. 2004, \aap, 416, L35

\bibitem[\protect\citeauthoryear{{Richard} et~al.}{{Richard}
  et~al.}{2006}]{Richard06}
{Richard}, J., {Pell{\'o}}, R., {Schaerer}, D., {Le Borgne}, J.-F.,  \&
  {Kneib}, J.-P. 2006, \aap, 456, 861

\bibitem[\protect\citeauthoryear{{Ricotti} \& {Shull}}{{Ricotti} \&
  {Shull}}{2000}]{Ricotti00}
{Ricotti}, M.,  \& {Shull}, J.~M. 2000, \apj, 542, 548

\bibitem[\protect\citeauthoryear{{Santos} et~al.}{{Santos}
  et~al.}{2004}]{Santos04}
{Santos}, M.~R., {Ellis}, R.~S., {Kneib}, J.-P., {Richard}, J.,  \& {Kuijken},
  K. 2004, \apj, 606, 683

\bibitem[\protect\citeauthoryear{{Schaerer}}{{Schaerer}}{2003}]{Schaerer03}
{Schaerer}, D. 2003, \aap, 397, 527

\bibitem[\protect\citeauthoryear{{Shapley} et~al.}{{Shapley}
  et~al.}{2003}]{Shapley03}
{Shapley}, A.~E., {Steidel}, C.~C., {Pettini}, M.,  \& {Adelberger}, K.~L.
  2003, \apj, 588, 65

\bibitem[\protect\citeauthoryear{{Shapley} et~al.}{{Shapley}
  et~al.}{2006}]{Shapley06}
{Shapley}, A.~E., {Steidel}, C.~C., {Pettini}, M., {Adelberger}, K.~L.,  \&
  {Erb}, D.~K. 2006, \apj, 651, 688

\bibitem[\protect\citeauthoryear{{Shimasaku} et~al.}{{Shimasaku}
  et~al.}{2006}]{Shimasaku06}
{Shimasaku}, K., et~al. 2006, \pasj, 58, 313

\bibitem[\protect\citeauthoryear{{Smith} et~al.}{{Smith}
  et~al.}{2005}]{Smith05}
{Smith}, G.~P., {Kneib}, J.-P., {Smail}, I., {Mazzotta}, P., {Ebeling}, H.,  \&
  {Czoske}, O. 2005, \mnras, 359, 417

\bibitem[\protect\citeauthoryear{{Smith} et~al.}{{Smith}
  et~al.}{2006}]{Smith06}
{Smith}, G.~P., {Sand}, D.~J., {Egami}, E., {Stern}, D.,  \& {Eisenhardt},
  P.~R. 2006, \apj, 636, 575

\bibitem[\protect\citeauthoryear{{Smith} et~al.}{{Smith}
  et~al.}{2002}]{Smith02}
{Smith}, G.~P., et~al. 2002, \mnras, 330, 1

\bibitem[\protect\citeauthoryear{{Spergel} et~al.}{{Spergel}
  et~al.}{2006}]{Sp06}
{Spergel}, D.~N., et~al. 2006, \apj,submitted

\bibitem[\protect\citeauthoryear{{Spergel} et~al.}{{Spergel}
  et~al.}{2003}]{Sp03}
{Spergel}, D.~N., et~al. 2003, \apjs, 148, 175

\bibitem[\protect\citeauthoryear{{Stanway} et~al.}{{Stanway}
  et~al.}{2004}]{Stanway04}
{Stanway}, E.~R., {Bunker}, A.~J., {McMahon}, R.~G., {Ellis}, R.~S., {Treu},
  T.,  \& {McCarthy}, P.~J. 2004, \apj, 607, 704

\bibitem[\protect\citeauthoryear{{Stark} et~al.}{{Stark}
  et~al.}{2006}]{Stark06}
{Stark}, D.~P., {Bunker}, A.~J., {Ellis}, R.~S., {Eyles}, L.~P.,  \& {Lacy}, M.
  2006, Accepted in ApJ, astro-ph/0604250

\bibitem[\protect\citeauthoryear{{Steidel}, {Pettini}, \&
  {Adelberger}}{{Steidel} et~al.}{2001}]{Steidel01}
{Steidel}, C.~C., {Pettini}, M.,  \& {Adelberger}, K.~L. 2001, \apj, 546, 665

\bibitem[\protect\citeauthoryear{{Stern} et~al.}{{Stern}
  et~al.}{2005}]{Stern05}
{Stern}, D., {Yost}, S.~A., {Eckart}, M.~E., {Harrison}, F.~A., {Helfand},
  D.~J., {Djorgovski}, S.~G., {Malhotra}, S.,  \& {Rhoads}, J.~E. 2005, \apj,
  619, 12

\bibitem[\protect\citeauthoryear{{Stiavelli}, {Fall}, \& {Panagia}}{{Stiavelli}
  et~al.}{2004}]{Sti04a}
{Stiavelli}, M., {Fall}, S.~M.,  \& {Panagia}, N. 2004, \apj, 600, 508

\bibitem[\protect\citeauthoryear{{Swinbank} et~al.}{{Swinbank}
  et~al.}{2006}]{Swinbank06}
{Swinbank}, A.~M., {Bower}, R.~G., {Smith}, G.~P., {Smail}, I., {Kneib}, J.-P.,
  {Ellis}, R.~S., {Stark}, D.~P.,  \& {Bunker}, A.~J. 2006, \mnras, 368, 1631

\bibitem[\protect\citeauthoryear{{Totani} et~al.}{{Totani}
  et~al.}{2006}]{Totani}
{Totani}, T., {Kawai}, N., {Kosugi}, G., {Aoki}, K., {Yamada}, T., {Iye}, M.,
  {Ohta}, K.,  \& {Hattori}, T. 2006, \pasj, 58, 485

\bibitem[\protect\citeauthoryear{{Tumlinson}, {Giroux}, \& {Shull}}{{Tumlinson}
  et~al.}{2001}]{Tumlinson01}
{Tumlinson}, J., {Giroux}, M.~L.,  \& {Shull}, J.~M. 2001, \apjl, 550, L1

\bibitem[\protect\citeauthoryear{{Tumlinson} \& {Shull}}{{Tumlinson} \&
  {Shull}}{2000}]{Tumlinson00}
{Tumlinson}, J.,  \& {Shull}, J.~M. 2000, \apjl, 528, L65

\bibitem[\protect\citeauthoryear{{Weatherley}, {Warren}, \&
  {Babbedge}}{{Weatherley} et~al.}{2004}]{Weatherley04}
{Weatherley}, S.~J., {Warren}, S.~J.,  \& {Babbedge}, T.~S.~R. 2004, \aap, 428,
  L29

\bibitem[\protect\citeauthoryear{{Willis} \& {Courbin}}{{Willis} \&
  {Courbin}}{2005}]{Willis05}
{Willis}, J.~P.,  \& {Courbin}, F. 2005, \mnras, 357, 1348

\bibitem[\protect\citeauthoryear{{Wyithe} \& {Loeb}}{{Wyithe} \&
  {Loeb}}{2006}]{Wyithe06}
{Wyithe}, J.~S.~B.,  \& {Loeb}, A. 2006, \nat, 441, 322

\bibitem[\protect\citeauthoryear{{Yan} et~al.}{{Yan} et~al.}{2006}]{Y06}
{Yan}, H., {Dickinson}, M., {Giavalisco}, M., {Stern}, D., {Eisenhardt},
  P.~R.~M.,  \& {Ferguson}, H.~C. 2006, \apj, 651, 24

\bibitem[\protect\citeauthoryear{{Yan} et~al.}{{Yan} et~al.}{2005}]{Y05}
{Yan}, H., et~al. 2005, \apj, 634, 109

\bibitem[\protect\citeauthoryear{{Yan} \& {Windhorst}}{{Yan} \&
  {Windhorst}}{2004}]{yw04}
{Yan}, H.,  \& {Windhorst}, R.~A. 2004, \apjl, 600, L1

\end{thebibliography}

\newpage

\renewcommand{\theequation}{A-\arabic{equation}}
\setcounter{equation}{0}
\section*{APPENDIX}

The observing sequence for narrowband surveys often involves 
alternating between two narrowband filters (e.g. A and B) with slightly 
different central wavelengths.  In each filter, the observations are 
divided into subexposures with integration times that are comparable to 
the variability timescale of the sky background (typically 5-10 minutes).  
A composite image 
is made of all of the subexposures in each filter, resulting 
in two ``subsurveys'' slightly offset in redshift space. Subtracting 
the two composite images removes continuum soures, thereby allowing line 
emitters to be identified. Alternatively, instead of a second narrowband 
filter, 
a broadband filter can be used to identify line emitters \citep{Willis05}. 
Using either approach, if integration time t$\rm_{int}$ is 
devoted to a narrowband survey, then the total integration devoted to each 
individual ``subsurvey'' is only (t$\rm_{int}$/2) in length. In the following 
calculations, we assume a total integration time of 60 hours.

The field of view of the survey camera is one of the key parameters 
in determining the survey sensitivity.  For a single exposure, the field 
of view is limited by the size of the detector and 
the plate scale of the detector.  The plate scale is typically set such 
that each pixel samples 1/4 of a spatial resolution element. 
Assuming seeing-limited 
observations with a 0\farcs6 seeing disk, this corresponds to 0.1-0.2
arcseconds per pixel.  Until very recently, only 1024$\times$1024 
pixel detectors were available on 8-10 meter telescopes in the near-infrared; 
given typical plate scales, this allows fields of view between 
3 and 12 arcmin$^2$.  Recently, larger format cameras have become 
available (e.g. MOIRCS on Subaru) allowing fields of view of $\simeq$30
arcmin$^2$ to be observed in single exposures.  In the signal to noise 
calculations that follow, we consider both a detector with a field 
of view of 4 arcmin$^2$ (similar to that of NIRI on Gemini) as 
well as one with a field of view of 30 arcmin$^2$.  

In the near-infrared, the noise is dominated by bright atmospheric 
emission lines from OH molecules; hence the background in a given 
exposure is dependent on the central wavelength and bandwidth of 
the narrowband filter.  The central wavelength is chosen to lie 
in $>$50\AA~ gaps between bright OH lines, where the atmospheric 
transmission is near 100\%.  We assume a central wavelength of 11905~\AA~ 
for our predictions. The signal to noise in a single exposure 
is minimized when the filter 
bandwidth is matched to the FWHM of the emission line, which is 
typically 5-10~\AA\ for Lyman-$\alpha$ emitters at high redshift 
(Shimasaku et al. 2006); hence, we adopt~10\AA~ as the FWHM for our 
calculations.  We also consider a filter 
with a FWHM of 100~\AA; while the noise in a single exposure is larger
for this filter, the redshift range covered is ten times larger 
($\Delta z$=0.1 vs. $\Delta z$=0.01). 

The received signal from the science source is estimated as follows,
\begin{equation}
S_{source} = \int f_{\lambda,source}\tau(\lambda)\,d\lambda~A_{tel}\eta_{tel} \eta_{inst} \eta_{QE} \eta_{atm} (t_{int}/2)
\end{equation}
where f$_{\lambda,source}$ is the spectrum of the source in units of 
photons s$^{-1}$ cm$^{-2}$ \AA$^{-1}$, $\tau(\lambda)$ is the filter transmission 
profile (assumed gaussian 
with FWHM defined above), $\eta_{tel}$ is the telescope optics throughput, 
$\eta_{inst}$ is the throughput of the instrument, $\eta_{QE}$ is the 
quantum efficiency of the detector, $\eta_{atm}$ is the atmospheric 
transmision, $A_{tel}$ is the collecting area of the telescope.
As discussed in $\S5.1$, we expect the line profile of high-redshift Lyman-$\alpha$ 
emitters to be asymmetric with the blue side of the line absorbed.  Along these lines, 
we assume the line profile is a half-Gaussian with no flux shortward 
of the central wavelength.  While this profile is perhaps oversimplified, 
the S/N calculations do not change signficantly if the profile is altered.
We adopt throughput and 
quantum efficiency values that are consistent with those expected 
for DAzLE ($\eta_{tel}$=0.6, $\eta_{inst}$=0.5, $\eta_{QE}$=0.7), 
a near infrared narrowband imager designed to find 
$z>7$ Lyman-$\alpha$ emitters \cite{Horton04}.  The atmospheric transmission, 
$\eta_{atm}$, is computed following models presented in \cite{Lord92} assuming 
an air mass of 1.5 and a 1.0 mm water vapour column \footnote{available from 
{\tt http://www.gemini.edu/sciops/ObsProcess/obsConstraints/atm-models/trans\_10\_15.dat}} 
and is near unity for the OH window considered.  We assume a telescope 
collecting area of 7.9$\times$10$^{5}$ cm$^2$ corresponding to the 
size of Keck.  The photometric aperture is matched to the area 
subtended under reasonably good seeing conditions ($\simeq$ 0\farcs6 
diameter seeing disk).  The integration time is only (t$_{int}$/2) because the total 
integration time is split between two different narrowband filters, as 
described above.

The noise in an exposure of time t$\rm_{int}$ is a function of the 
integrated source counts, sky background, detector dark current, and 
detector read noise.  Formally, the noise is defined as 
\begin{equation}
N = \sqrt{N_{sky} + N_{source} + N_{DC} + RN^2}
\end{equation}
where $N_{sky}$, N$_{source}$, N$_{DC}$ are the number of 
electrons from the sky background, science source, and dark current, and RN 
is the read noise. 

We compute the expected background count rate utilizing the Gemini model 
for the near-infrared sky background spectrum\footnote{available from 
{\tt http://www.gemini.edu/sciops/ObsProcess/obsConstraints/atm-models/nearIR\_skybg\_16\_15.dat}}.  
The model inludes the contribution from 
atmospheric OH emission lines, zodiacal emission (from a 5800 
K blackbody), and thermal emission from the atmosphere (from a 
250 K blackbody).  The total counts
from the sky background in a photometric aperture of area a$_{phot}$
in the composite image, is given by 
\begin{equation}
N_{sky} = \int f_{\lambda,sky}~\tau(\lambda)\,d\lambda~\eta_{tel}\eta_{inst}\eta_{QE}~A_{tel}~a_{phot}~(t_{int}/2)
\end{equation}
where $f_{\lambda,sky}$ is the sky background in photons s$^{-1}$
cm$^{-2}$ \AA$^{-1}$ arcsec$^{-2}$, $\tau$ is the filter transmission 
profile (assumed gaussian with FWHM defined above), a$_{phot}$ has units of 
arcsec$^2$, and all other parameters are the same as defined above.

The total counts associated with the dark current is given by 
\begin{equation}
N_{DC} = n_{pix} dc (t_{int}/2) 
\end{equation}
where n$_{pix}$ is the number of pixels in the photometric aperture area 
and dc is the dark current in units of electrons per second per pixel.  
The number of pixels in the aperture is derived assuming a plate 
scale of 0\farcs15.  The typical dark 
current for near-IR imagers on 8-10 meter telescopes is 0.25 electrons 
s$^{-1}$ pixel$^{-1}$; hence we adopt this value in our calculations.

The read noise contribution to the total noise is computed assuming 
Fowler sampling is used, which reduces the read noise by a factor 
of $\sqrt 4/n_{read}$ where n$_{read}$ is the number of reads at the 
beginning and end of each exposure.  The total read noise in an 
aperture of size n$\rm_{pix}$ in the composite images is thus 
given by 
\begin{equation}
RN=rn \sqrt \frac{4~n_{pix} n_{exp}}{n_{read}} 
\end{equation}
where rn is the detector read noise (for single readout) in units of electrons 
per pixel and n$_{exp}$ is the number of exposures in the composite 
frame.  We adopt a read noise of 10 electrons per pixel \citep{Horton04}
for a single readout and assume the detector is read out 16 times per 
exposure.

The 5$\sigma$ limiting flux is defined as the line flux for which 
the signal from the science source is a factor of five greater 
than the noise.  For the instrument properties assumed, the 5$\sigma$ 
limiting flux in a 10 hour narrowband survey is 3-9$\times$10$^{-18}$ 
erg cm$^{-2}$ s$^{-1}$ 
for a 10\AA~ and 100\AA~ narrowband filter, respectively.  
At z=9, this corresponds to a Lyman-$\alpha$ luminosity of 
0.6-2$\times$10$^{42}$ erg s$^{-1}$ assuming 
a magnification of $\times$5.  The survey volume is computed 
following equations 3 and 4 
presented in $\S$5.1.2 and assuming a magnification distribution 
for one of the most well-studied clusters, Abell 1689.

We also compute the efficiency of a conventional narrowband survey.  
The survey parameters are identical to those assumed for the lensing 
survey.  The only difference is that in computing the limiting 
luminosity and survey volume, we do not fold in the magnification 
due to the lensing cluster.

The parameters chosen for the spectroscopic survey correspond to 
the properties of NIRSPEC on Keck.  The limiting sensitivity in a 
spectroscopic survey is computed in the same way as for the 
narrowband survey.  The spectroscopic survey strategy is described in $\S3.2$.  
We divide the total integration time into many 1.5 hour observations; 
each 1.5 hour integration is further divided into nine 10 minute exposures.
One key difference between the spectroscopic and narrowband survey is that 
the entire integration time is devoted to a single survey; hence, 
the (t$_{int}$/2) in the equations above become t$_{int}$ for 
spectroscopic surveys.  

Most of the parameters used in the spectroscopic S/N computation are 
identical to those used above.  The read noise in each ten minute 
exposure is 25 electrons pixel$^{-1}$ assuming 16 total reads; while 
slightly higher than in the narrowband survey described above, the 
noise is still sky-background dominated.

The median 5$\sigma$ limiting flux for the spectroscopic survey 
is 8$\times$10$^{-18}$
erg cm$^{-2}$ s$^{-1}$.  We assume the longslit is placed along 
the critical lines of the lensing clusters listed in Table 1, which 
results in a significantly higher median magnification than in the 
narrowband survey.  For a magnification of $\times$25, this corresponds 
to a limiting Lyman-$\alpha$ luminosity of 4$\times$10$^{41}$ erg s$^{-1}$.  

\begin{deluxetable}{lcccc}
\tablewidth{0pt}
\tablecaption{Clusters surveyed}
\tablehead{
Cluster & 
Redshift & 
RA\tablenotemark{a} & 
Dec\tablenotemark{b} & 
Lens Model Reference} 
\startdata
Cl 0024 & $0.39$ & 00 26 35.5  &  +17 09 50.7  &  (1)  \\

Abell 68 & $0.255$ & 00 37 06.8  & +09 09 23.4 & (2)\\

Abell 370 & $0.375$ & 02 39 53.1  & -01 34 54.8 & (3)\\

MS 0451 & $ 0.55 $ & 04 54 10.6  & -03 00 50.7 &  (4)  \\

Abell 963 & $0.206$ & 10 17 03.7  & +39 02 49.2 & (2) \\

Abell 1689 & $0.183$ & 13 11 29.4  & -01 20 28.7 & (5) \\

Abell 2218 & $0.176$ & 16 35 49.3  & +66 12 43.5 & (6) \\

Abell 2219 & $0.226$ & 16 40 19.8  & +46 42 41.9 & (2) \\

Abell 2390 & $0.228$ & 21 53 36.9  & +17 41 43.4 & (7) \\
\enddata
\tablenotetext{a}{units of HH MM SS}
\tablenotetext{b}{units of +DD MM}
\tablerefs{(1) \cite{Kneib03}; (2) \cite{Smith05}
(3) \cite{Bez99}; (4) \cite{Borys04}, (5) Limousin et al. 
{\em in prep}; (6) \cite{Kneib04}; (7)  \cite{Swinbank06}}
\end{deluxetable}

\begin{deluxetable}{llcc}
\tablewidth{0pt}
\tablecaption{NIRSPEC Survey Observations}
\tablehead{
\colhead{Date} & \colhead{Cluster} & \colhead{Position Angle\tablenotemark{a}} & 
\colhead{Integration time\tablenotemark{b}} }
\startdata
Aug 2004  & Abell 2390 1  & 117     & 12.0  \\
          & Abell 2219 1  & 119.2   & 9.6   \\
          & Abell 68 1    &  330.   & 13.2  \\
Jan 2005  & MS 0451 1     & 120.9   & 11.4  \\
          & Abell 963 1   & 177.2   & 10.8  \\
          & Abell 963 2   & 175.0   & 5.4   \\
          & Abell 1689 1  & 43      & 5.4    \\
June 2005  & Abell 1689 1   & 43    & 3.6    \\
           & Abell 1689 2   & 205   & 10.8   \\
           & Abell 1689 3   & 139.1 & 10.8   \\
           & Abell 2218 1   & 153   & 10.8   \\
           & Abell 2218 2   & 134   & 5.4    \\
           & Abell 2219 1   & 119.2 & 5.4    \\
           & Abell 2219 2   & 160   & 16.2   \\
           & Abell 2390 2   & 109   & 10.2   \\
           & Abell 2390 3   & 20.0  & 6      \\
         
Oct 2005  & Abell 68 2     & 300    &  10.8   \\ 
          & Cl 0024 1      & 139.7  &  5.4   \\
          & Cl 0024 2      & 160.2  &  4.8   \\
          & MS 0451 2      & 100.1  &  7.2  \\
          & Abell 2390 3   & 20.0   &  5.4   \\
          & Abell 2390 4   & 90.0   &  4.8   \\
          & Abell 2219 2   & 160    & 2.4    \\
\enddata
\tablenotetext{a}{in degrees North through East}
\tablenotetext{b}{in ksec}
\end{deluxetable}

\begin{deluxetable}{lccccccc}
\tablewidth{0pt}
\tabletypesize{\fontsize{7}{10}\selectfont}
\tablecaption{NIRSPEC \lya~ Candidates\label{candidates}}
\tablehead{
\colhead{Candidate}      & 
\colhead{RA(J2000)}      &
\colhead{DEC(J2000)}     &
\colhead{flux}           & 
\colhead{$\lambda$}      & 
\colhead{z$_{\rm~Ly\alpha}$} &
\colhead{Log$_{10}$ L$_{\rm~Ly\alpha}$} & 
\colhead{Log$_{10}$ Mag}             \\
\colhead{}            & 
\colhead{}            &
\colhead{}            &
\colhead{($\ergscm$)} & 
\colhead{($\mu$m)}    & 
\colhead{}            & 
\colhead{($\ergs$)}   & 
\colhead{}}            
\startdata 
A68c1   & 00 37 06.10 & $+$09 09 18.7 & 2.2$\pm0.3\times$10$^{-17}$ & 1.254 & 9.32  & 41.2$^{+0.1}_{-0.7}  $ &  2.2$^{+0.7}_{-0.1}$   \\
A1689c1 & 13 11 31.85 & $-$01 20 40.0 & 2.3$\pm0.3\times$10$^{-17}$ & 1.366 & 10.23 & 42.3$^{+0.02}_{-0.01}$ &  1.2$^{+0.01}_{-0.02}$ \\
A1689c2 & 13 11 31.26 & $-$01 20 29.8 & 4.0$\pm0.7\times$10$^{-17}$ & 1.173 & 8.65  & 42.7$^{+0.07}_{-0.08}$ &  0.8$^{+0.01}_{-0.03}$  \\
A1689c3 & 13 11 28.43 & $-$01 19 44.5 & 4.1$\pm0.6\times$10$^{-17}$ & 1.259 & 9.35  & 42.1$^{+0.06}_{-0.08}$ &  1.5$^{+0.04}_{-0.03}$  \\
A2219c1 & 16 40 23.64 & $+$46 42 26.5 & 4.8$\pm0.6\times$10$^{-17}$ & 1.215 & 8.99  & 41.8$^{+0.09}_{-0.07}$ &  1.9$^{+0.04}_{-0.07}$  \\  
A2219c2 & 16 40 23.04  &$+$46 42 43.2 & 2.1$\pm0.3\times$10$^{-17}$ & 1.209 & 8.94  & 42.0$^{+0.06}_{-0.09}$ &  1.3$^{+0.06}_{-0.03}$   \\
\enddata
\end{deluxetable}

\begin{deluxetable}{llccccc}
\tablewidth{0pt}
\tabletypesize{\fontsize{9}{10}\selectfont}
\tablecaption{Wavelengths of Additional Emission Lines for Low-z Scenarios}
\tablehead{
\colhead{Line}      & 
\colhead{z}      &
\colhead{$\lambda_{\rm~Ly\alpha}$}     &
\colhead{$\lambda_{\rm~[OII]}$}           & 
\colhead{$\lambda_{\rm~H\beta}$} &
\colhead{$\lambda_{\rm~[OIII]}$} &
\colhead{$\lambda_{\rm~H\alpha}$} 
\\
\colhead{}      & 
\colhead{}      &
\colhead{($\mu$m)}     &
\colhead{($\mu$m)}           & 
\colhead{($\mu$m)} &
\colhead{($\mu$m)} &
\colhead{($\mu$m)}}

\startdata
H$\alpha$   & 0.85 &  0.2253  &  0.6905\tablenotemark{a}    & 0.9007 & 0.9188/0.9277 & 1.2160 \\   
$\rm~[OIII]$& 1.43\tablenotemark{b} &  0.2953   & 0.9051 & 1.1805 &  1.2043/1.2160 & 1.5939 \\  
            & 1.45\tablenotemark{c} &  0.2982   & 0.9139 & 1.1920 &  1.2160/1.2278 &  1.6093   \\  
H$\beta$    & 1.50 &  0.3042  & 0.9323 & 1.2160 &  1.2405/1.2525 & 1.6418  \\  
$\rm~[OII]$ & 2.26 &  0.3967  & 1.2160 & 1.5860 &  1.6180/1.6336 & 2.1413  \\  
Ly$\alpha$  & 9.0  &  1.2160  & 3.7270 & 4.8610 &  4.9590/5.0070 & 6.5630  \\
\enddata
\tablenotetext{a}{Throughout the table, we take the luminosity-weighted value of [OII], $\lambda$=3727 \AA, for conciseness.}
\tablenotetext{b}{$\lambda$=5007 \AA}
\tablenotetext{c}{$\lambda$=4959 \AA}

\end{deluxetable}

\begin{deluxetable}{cllllll}
\tablewidth{0pt}
\tabletypesize{\fontsize{7}{10}\selectfont}
\tablecaption{Constraints on Low-redshift Interpretations of Candidates}
\tablehead{
\colhead{Line}      & 
\colhead{z}      &
\colhead{$\rm~f_{\rm~Ly\alpha}$}     &
\colhead{$\rm~f_{\rm~[OII]}$}           & 
\colhead{$\rm~f_{\rm~H\beta}$} &
\colhead{$\rm~f_{\rm~[OIII]_1}$/$\rm~f_{\rm~[OIII]_2}$} &
\colhead{$\rm~f_{\rm~H\alpha}$} 
\\
\colhead{}      & 
\colhead{}      &
\colhead{(10$^{-17}$ erg cm$^{-2}$ s$^{-1}$)}     &
\colhead{(10$^{-17}$ erg cm$^{-2}$ s$^{-1}$)}           & 
\colhead{(10$^{-17}$ erg cm$^{-2}$ s$^{-1}$)} &
\colhead{(10$^{-17}$ erg cm$^{-2}$ s$^{-1}$)} &
\colhead{(10$^{-17}$ erg cm$^{-2}$ s$^{-1}$)}}
\startdata
\tableline\\*
\multicolumn{7}{c}{Abell 68 c1}\\*
\tableline\\*
H$\alpha$      & 0.91 & N/A       & 0.63-2.4 (0.07) & 0.79 (0.07)    &  N/A   / N/A                      & N/A \\   
$\rm~[OIII]_2$ & 1.51 & N/A       & N/A        & 0.31-2.8 (1)  & 0.73 (0.9)/ NA               & 0.4-13(2)  \\  
$\rm~[OIII]_1$ & 1.53 & N/A       & N/A        & 1.1-7.7  (6)  & NA / 6.9 (5)                 & 3.1-21 (2)  \\  
H$\beta$       & 1.58 & N/A       & N/A        & NA            & 0.66-4.4 (5) / 1.8-15 (3)     & 6.2 (2) \\  
$\rm~[OII]$    & 2.37 & 31 (0.02) & N/A        & 1.4 (5)       & 0.22-7.3 (7) / 2.2-5.1 (5)   & N/A      \\  
\tableline\\
\multicolumn{7}{c}{Abell 1689 c1}\\
\tableline\\
H$\alpha$      & 1.06 &  N/A      &  N/A       & N/A           &  N/A / N/A                      & N/A   \\   
$\rm~[OIII]_2$ & 1.71 &  N/A      &  N/A       & 0.33-2.9(6)    & 0.77(1) / N/A             & N/A     \\  
$\rm~[OIII]_1$ & 1.72 &  N/A      &  N/A       & 1.2-7.7(0.9)  & N/A  / 6.9(4)            & N/A     \\  
H$\beta$       & 1.79 &  N/A      &  N/A       & N/A           & 0.69-4.6(20) / 1.8-16(100)     & N/A     \\  
$\rm~[OII]$    & 2.64 &  N/A      &  N/A       & N/A           &  N/A / N/A                & N/A     \\  
\tableline\\
\multicolumn{7}{c}{Abell 1689 c2}\\
\tableline\\
H$\alpha$      & 0.78 &  N/A      &  N/A       & N/A           &  N/A  / N/A                    & N/A  \\   
$\rm~[OIII]_2$ & 1.33 &  N/A      &  N/A       & N/A           & 1.3 (5) / N/A                  & N/A     \\  
$\rm~[OIII]_1$ & 1.36 &  N/A      &  N/A       & N/A           & N/A     / 12(3)              & N/A     \\  
H$\beta$       & 1.40 &  N/A      &  N/A       & N/A           & 1.2-8.0(1) / 3.2-38(4) & N/A     \\  
$\rm~[OII]$    & 2.13 &  N/A      &  N/A       & N/A           & N/A   / N/A                         & N/A     \\  
\tableline\\
\multicolumn{7}{c}{Abell 1689 c3}\\
\tableline\\
H$\alpha$      & 0.94 & N/A      &   N/A       & N/A              & N/A  / N/A                          & N/A  \\   
$\rm~[OIII]_2$ & 1.54 & N/A      &   N/A       & 0.59-5.1 (8)     & 1.4 (2) / N/A                  & N/A     \\  
$\rm~[OIII]_1$ & 1.57 & N/A      &   N/A       & 6.2-11  (2)      & N/A    / 12 (5)                & N/A  \\  
H$\beta$       & 1.62 & N/A      &   N/A       & N/A              & 1.2-8.2 (3) / 3.3-29 (1) & N/A     \\  
$\rm~[OII]$    & 2.42 & N/A      &   N/A       & N/A              &  N/A  / N/A                        & N/A     \\  
\tableline\\
\multicolumn{7}{c}{Abell 2219 c1}\\
\tableline\\
H$\alpha$      & 0.85 & N/A     & 1.4-5.1 (0.3)   & 1.7 (1)           & 0.51-3.4 (0.4) / N/A              & N/A   \\   
$\rm~[OIII]_2$ & 1.43 & N/A     & N/A             & 0.69-6.0(2)       & 1.6 (5) / N/A                     & 1.9-17(2) \\  
$\rm~[OIII]_1$ & 1.45 & N/A     & N/A             & 2.4-16(0.9)       & N/A / 14(1)                       & 6.7-45 (7)  \\  
H$\beta$       & 1.50 & N/A     & N/A             & N/A               & 1.4-96 (2) / 3.8-34(0.9)          & 13 (3)        \\  
$\rm~[OII]$    & 2.26 & 35-130 (0.4)& N/A     & 1.6-6.0 (8)           & 1.8-3.2 (2) / 4.8-11 (5)          & N/A         \\  
\tableline\\
\multicolumn{7}{c}{Abell 2219 c2}\\
\tableline\\
H$\alpha$      & 0.84 & N/A        & 0.6-2.3 (0.4)   & 0.75 (1)         &  0.23-1.5(0.4) / 0.60-5.3 (0.5)    & N/A \\   
$\rm~[OIII]_2$ & 1.41 & N/A        & N/A         & 0.30-2.7 (2)         & 0.7 (3) / N/A                      & 0.84-7.3 (10)  \\  
$\rm~[OIII]_1$ & 1.44 & N/A        & N/A         & 0.40$-$13 (1)        & N/A / 6.3 (2)                      & 0.4-13 (1)  \\  
H$\beta$       & 1.49 & N/A        & N/A         & N/A                  & 0.63-4.2(3) / 1.7-15(0.8)          & 5.9(4) \\  
$\rm~[OII]$    & 2.24 & 15-58 (0.4)   & N/A      & 0.70-2.6 (2)         & 0.79-2.4(4) / 2.1-4.9 (5)          & N/A  \\  
\enddata
\tablecomments{The fluxes are predictions given the flux of the J-band 
feature and typical flux ratios for HII regions.  The value in 
paranthesis is the limiting line flux at the corresponding wavelength (see $\S$3.2 for details 
on how the limiting flux is computed). [OIII]$_1$ is $\lambda$4959\AA and [OIII]$_2$ is $\lambda$5007\AA}
\end{deluxetable}

\newpage

\begin{figure}
\figurenum{1}
\epsscale{.85}
\plotone{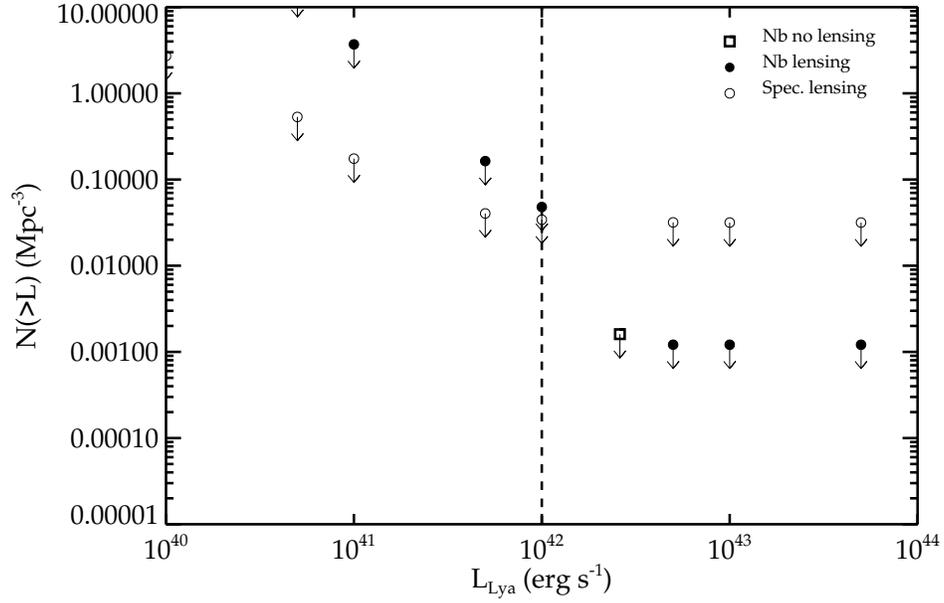}
\caption{Comparison of different survey strategies for locating
 Lyman-$\alpha$ emitters with current instrumentation in seeing-limited conditions.  
Solid circles represent 
the minimum source abundance that a 60 hour spectroscopic lensing observing 
campaign (over 10 clusters) could constrain (at 5$\sigma$); while open circles 
correspond to the minimum source abundance that a narrowband imaging lensing survey 
could constrain in the same exposure time.  Both survey techniques 
probe orders of magnitude fainter than a conventional narrowband 
survey (open square), albeit over much smaller volumes. 
At luminosities to the left of the dashed vertical line,
the spectroscopic approach covers a larger volume, and thus it 
is significantly more efficient than narrowband imaging
in constraining the abundance of the faintest ($\lsim$ 1 M$_\odot$
yr$^{-1}$) Lyman-$\alpha$ emitters. 
}
\label{}
\end{figure}

\begin{figure}
\figurenum{2}
\plotone{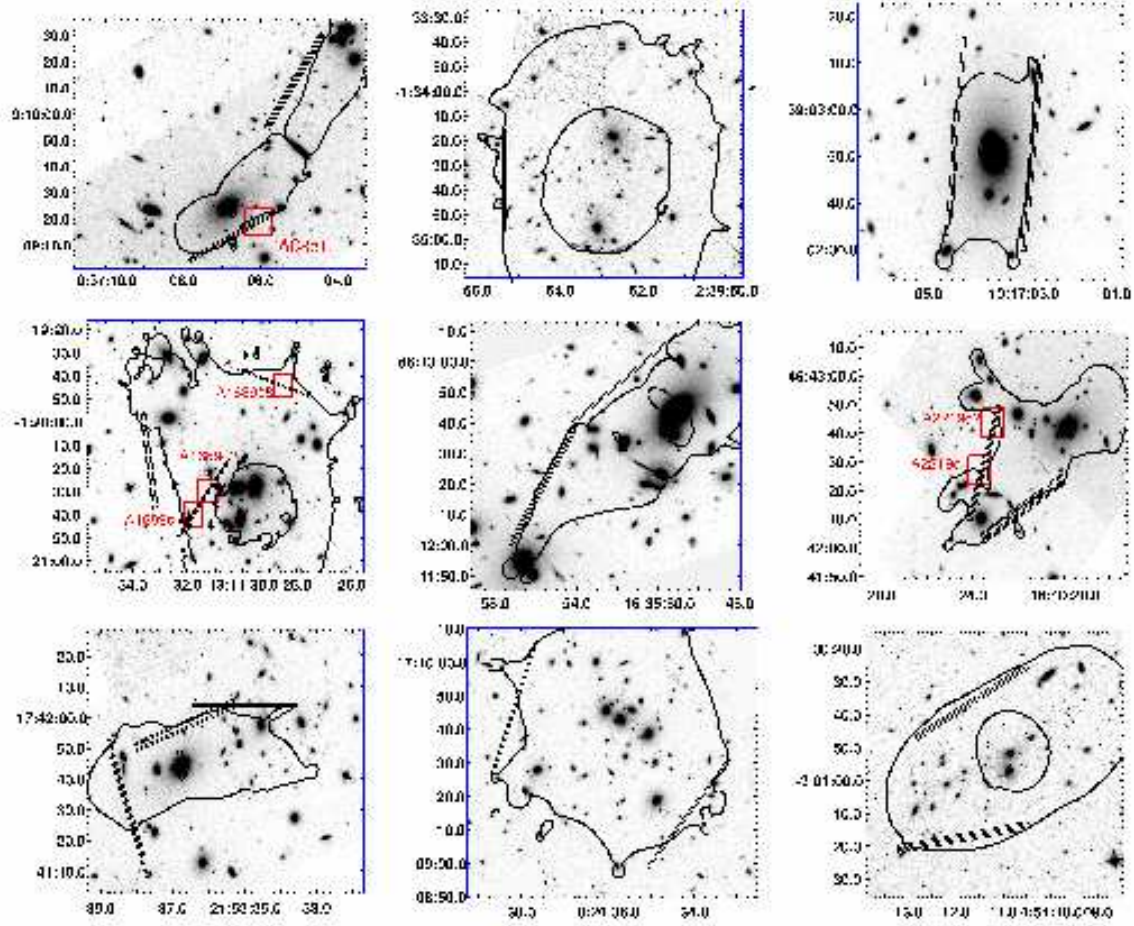}
  \caption{Survey clusters, with survey area and lensing critical
  curves. For each cluster, superposed on the \textit{HST}/WFPC2 image are 
  slit positions observed with NIRSPEC and the
  critical lines for a source at $z=8$ (dotted lines).  The solid
  curves bound areas where the magnification provided to $z=8$ 
  background sources exceeds
  a factor of 10.  From left to right and top to bottom are 
  Abell 68, Abell 370, Abell 963, Abell 1689, Abell 2218, Abell 2219,
  Abell 2390, CL 0024, MS 0451.}
  \label{fig:johan1}
\end{figure}

\begin{figure}
\figurenum{3}
\epsscale{.85}
\plotone{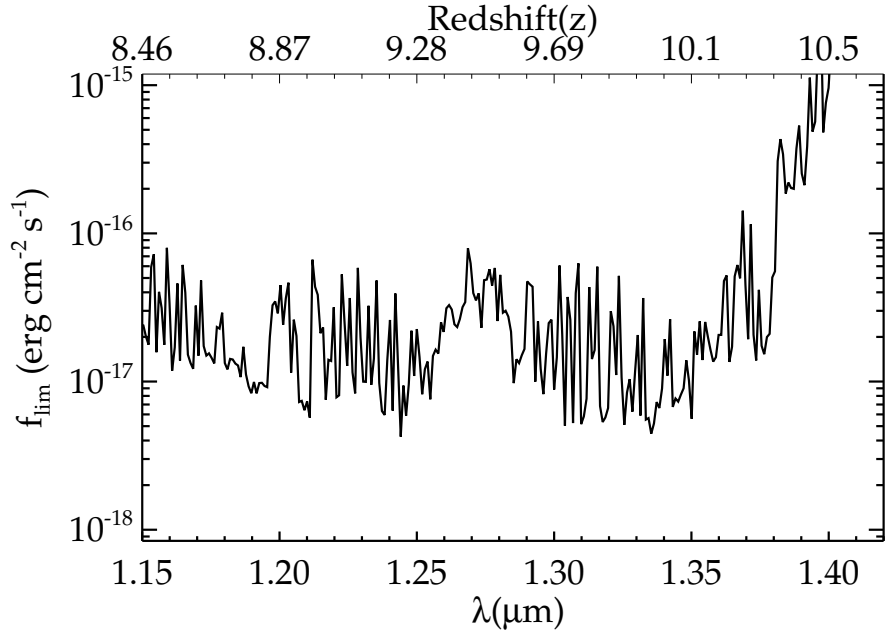}
\caption{Limiting flux for detection of 5$\sigma$ Lyman-$\alpha$ 
emission line for a typical NIRSPEC $J$ band exposure.  
The top axis denotes the redshift corresponding
to observed Lyman-$\alpha$ at the wavelength along the bottom axis.
{\bf OH bands lie throughout the J-band spectrum, signficantly increasing the 
limiting flux at those wavelengths.  The widest band occurs in the 
middle of the spectrum at $\simeq 1.26-1.28 \mu m$.  }
}
\label{}
\end{figure}

\begin{figure}
\figurenum{4}
\epsscale{1.2}
\plotone{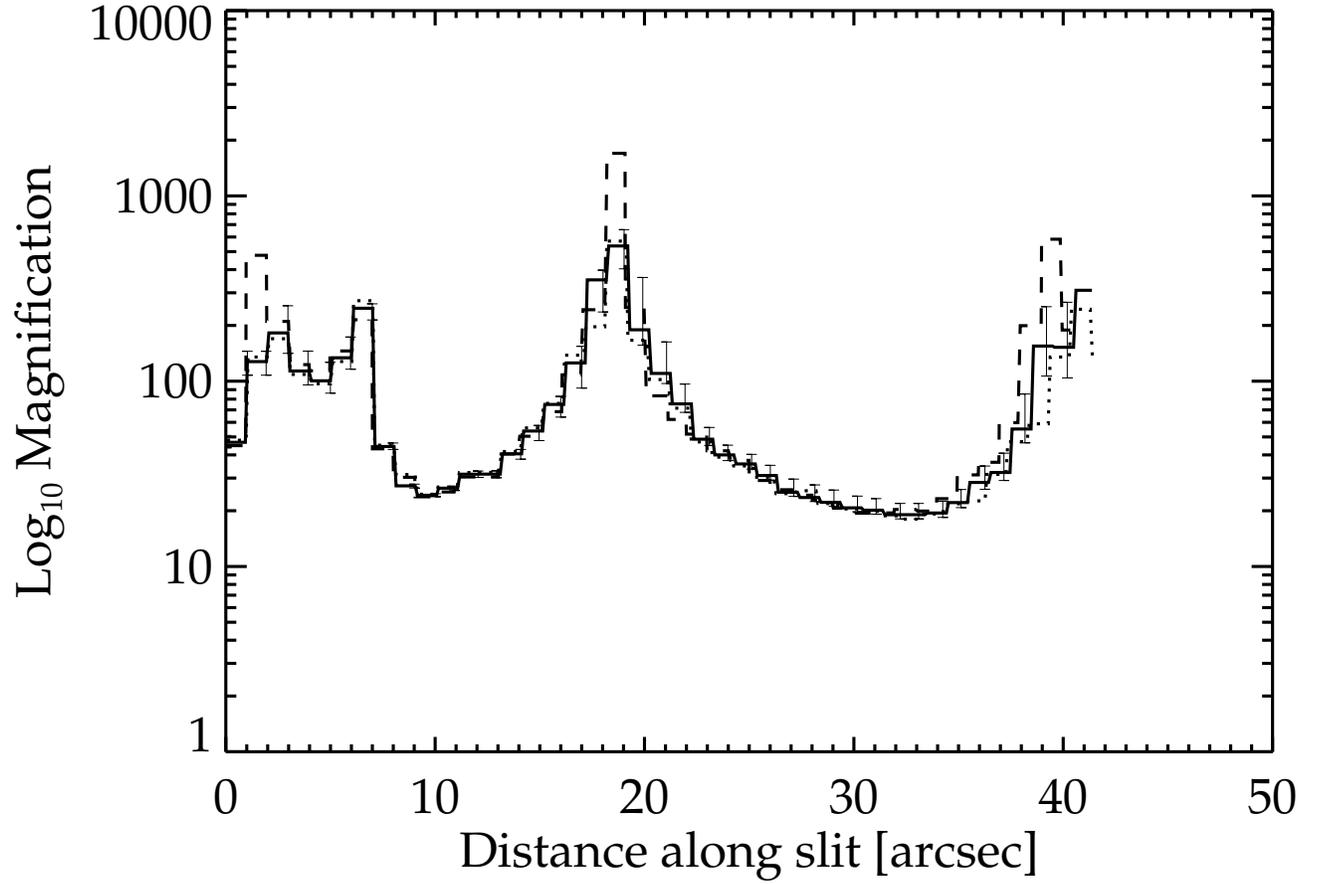}
\caption{Lensing magnification provided to background sources 
as a function of NIRSPEC slit position and redshift. This figure shows
the amplification at one of the slit positions 
along the critical line of Abell 2219.  The solid line corresponds
to the amplification provided to background sources at
z=9 whereas the dotted and dashed line denotes that provided to
sources at z=8.5 and 10.0, respectively.  The error in lensing magnification 
is plotted as a function of NIRSPEC slit position for sources at z=9.  The error 
does not vary significantly as a function of source redshift.  
Even allowing for uncertainty in the mass model, the magnification 
is still very high across the entire slit.
}
\label{}
\end{figure}

\begin{figure}
\figurenum{5}
\epsscale{.85}
\plotone{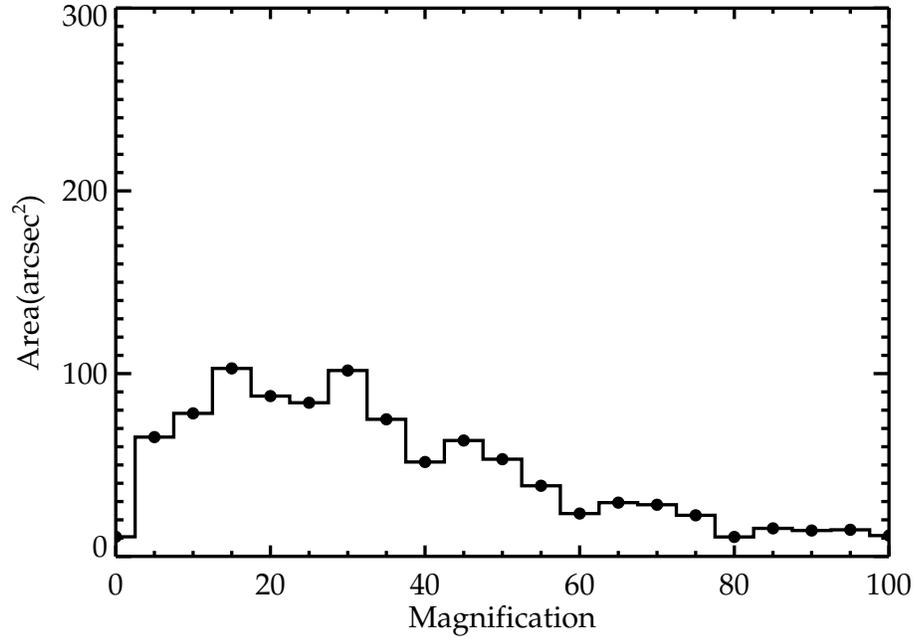}
\caption{Distribution of lensing magnfication across our total survey area.  
The median magnification across our survey area is $\simeq$20,  about 
a factor of 2 larger than in a similar optical spectroscopic 
lensing survey \citep{Santos04} {\bf (see \S3.4 for explanation)}.  A 
non-negligible fraction of each slit is magnified by larger than this factor.
}
\label{}
\end{figure}

\begin{figure*}
\figurenum{6}
\begin{center}
\includegraphics[width=7cm]{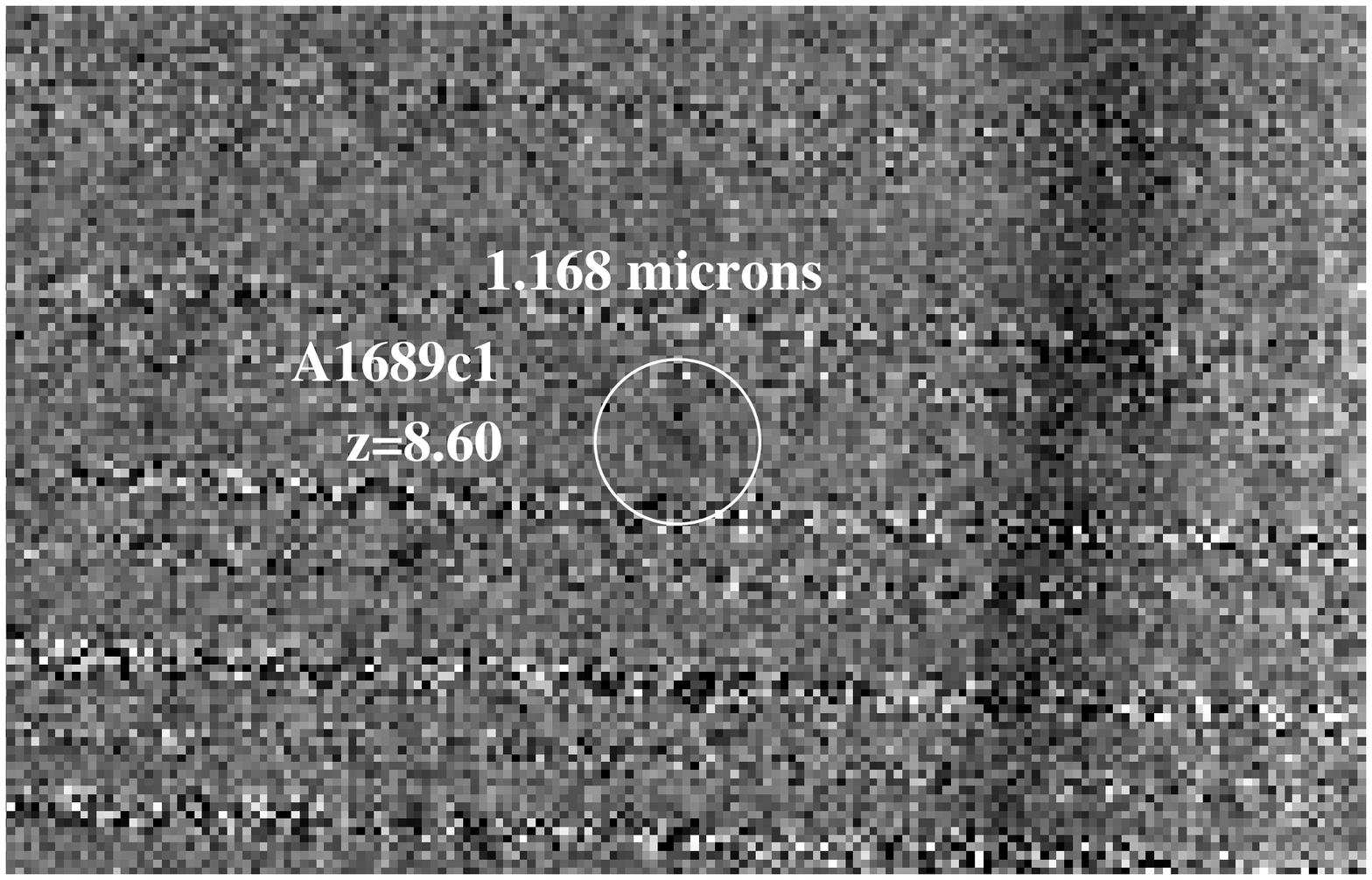}\includegraphics[width=7cm]{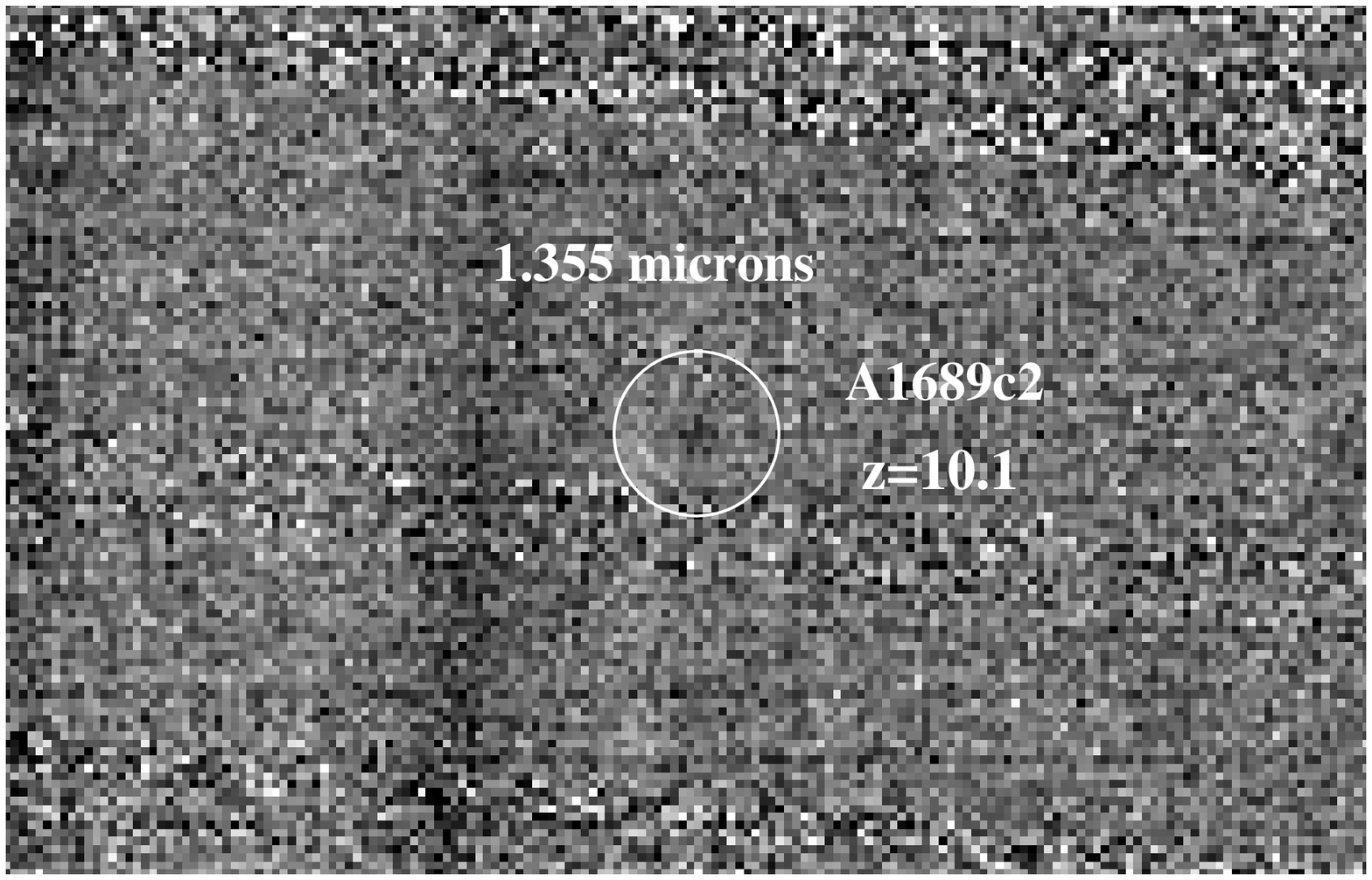}
\includegraphics[width=7cm]{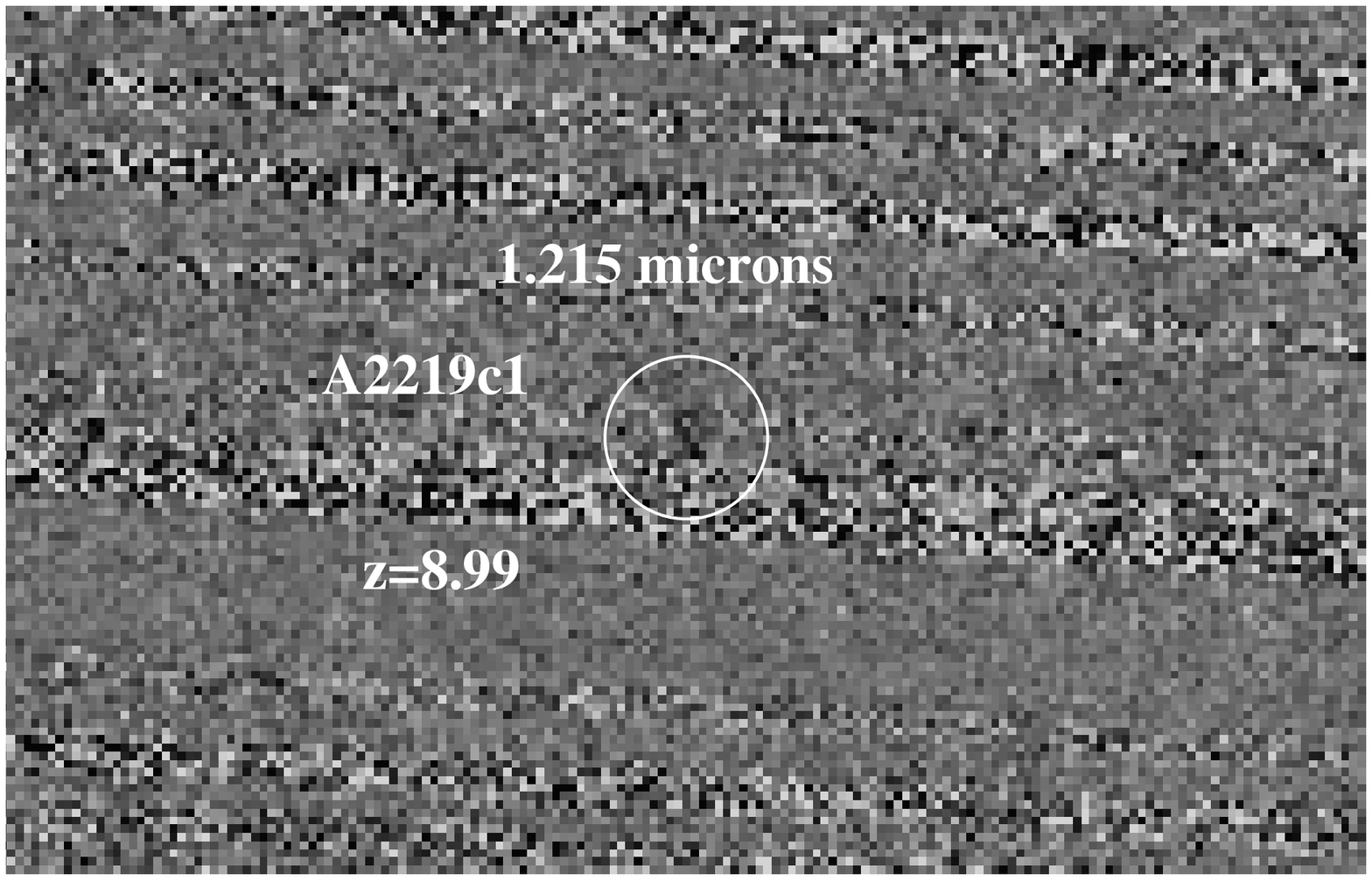}\includegraphics[width=7cm]{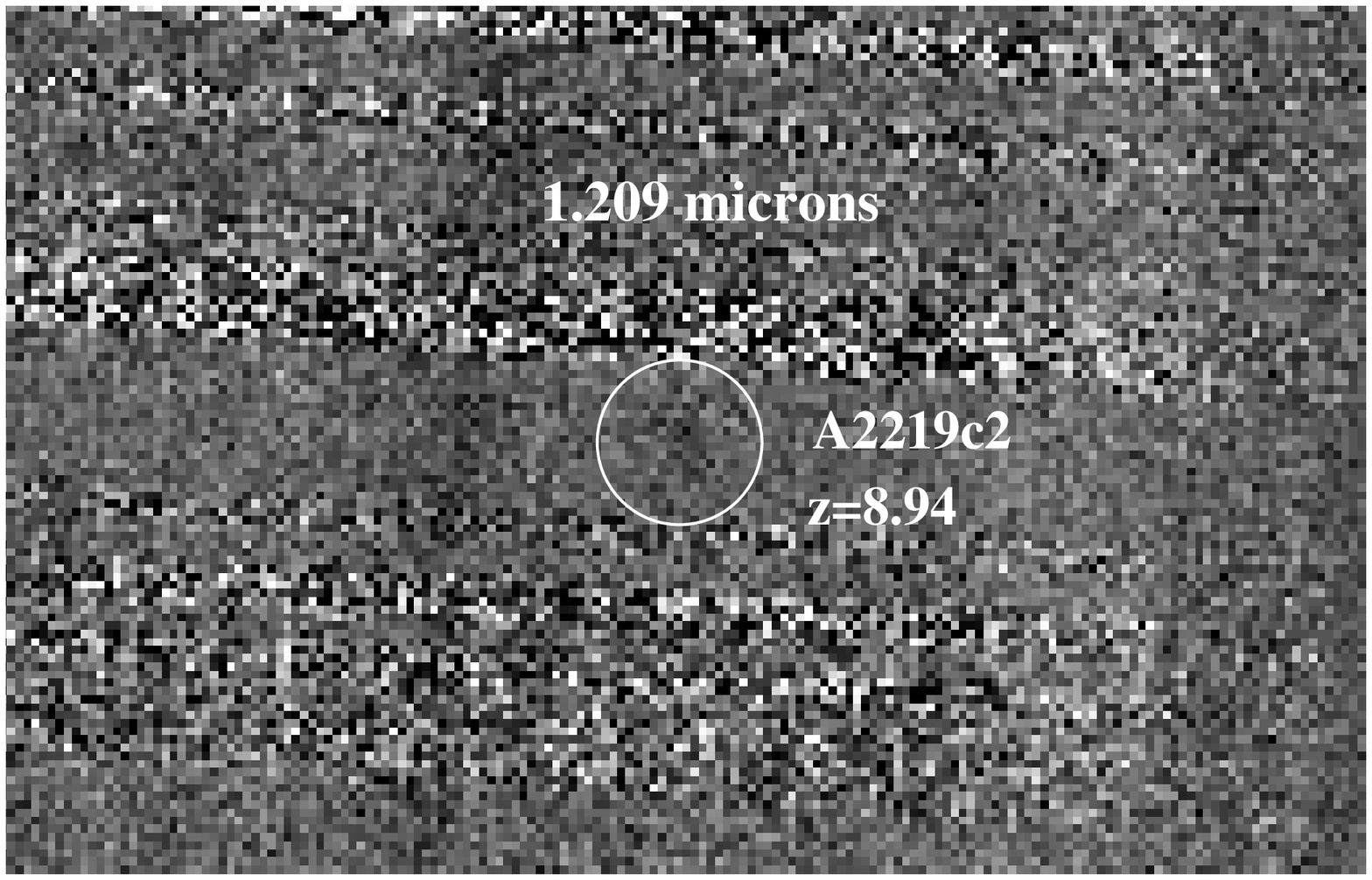}
\includegraphics[width=7cm]{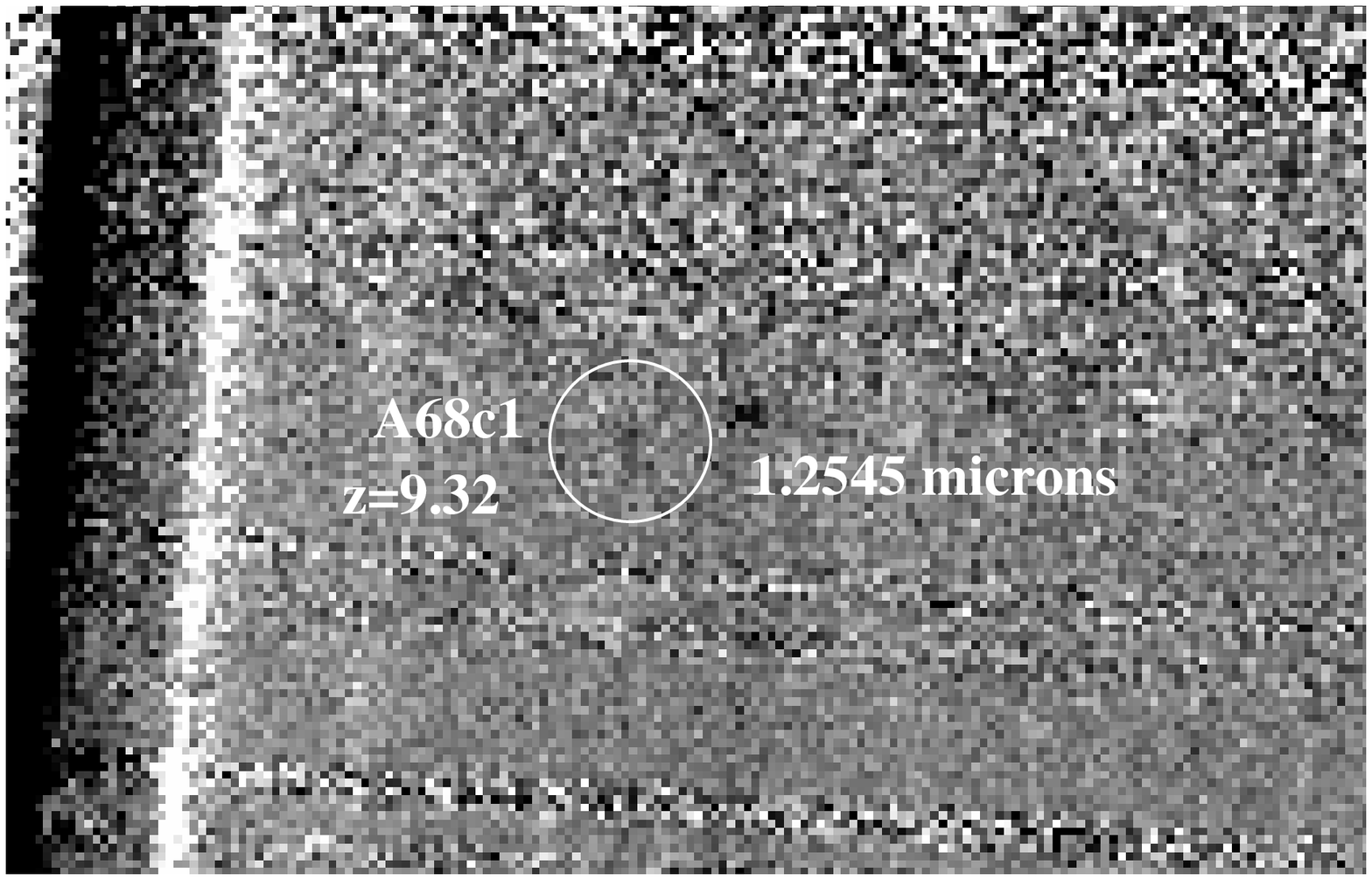}\includegraphics[width=7cm]{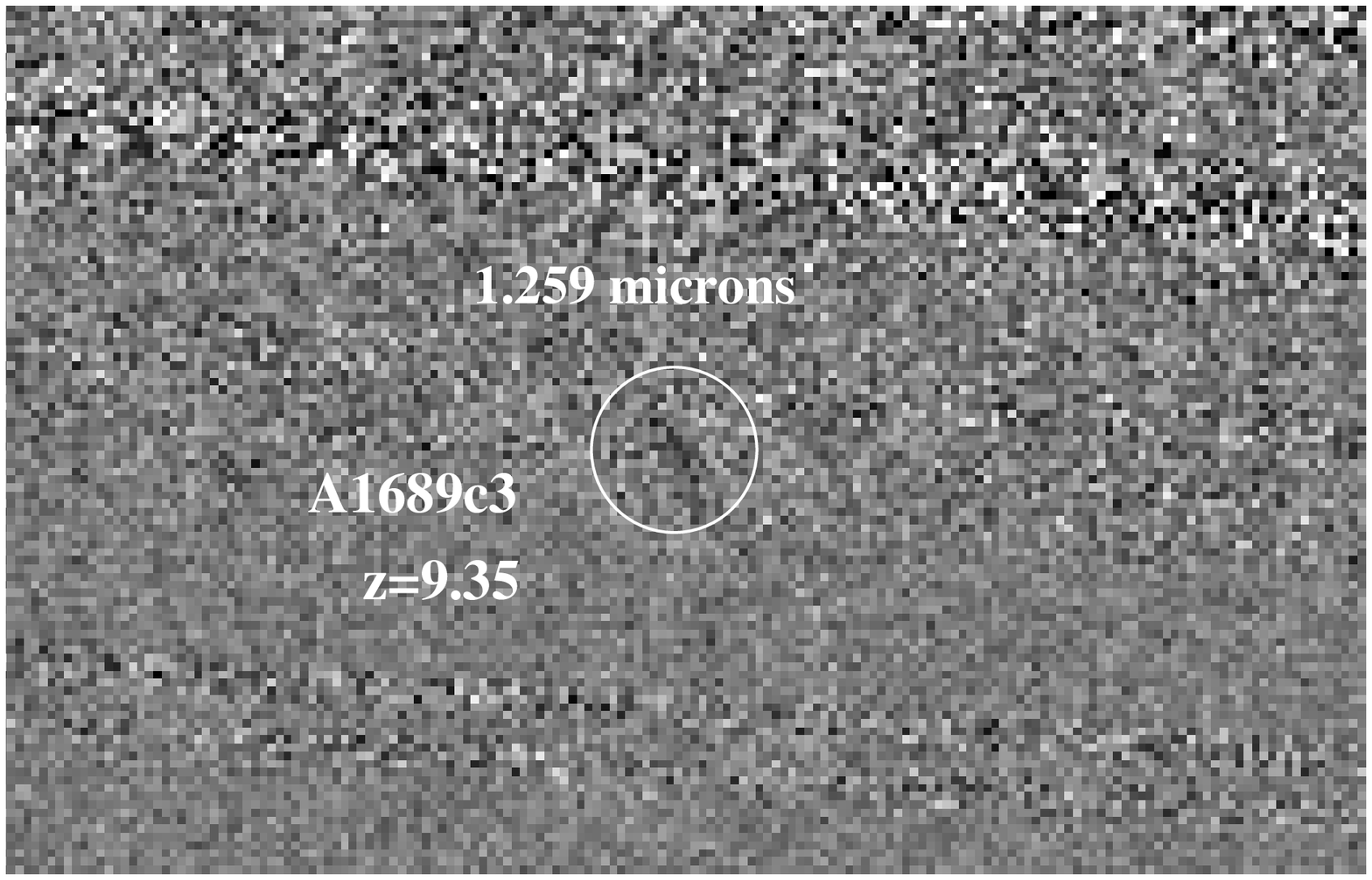}
\end{center}
\caption{Candidate Lyman-$\alpha$ emitters, each undetected in 
deep optical HST data.  If Lyman-$\alpha$, 
the emission line redshifts range between $z=8.7$ and $z=10.2$. Line 
fluxes range between $2-5\times$10$^{-17}$ erg cm$^{-2}$ s$^{-1}$ implying 
intrinsic (unlensed) star formation rates of 0.2-5 M$_\odot$ yr$^{-1}$, 
for nominal source assumptions. }
\end{figure*}

\begin{figure}
\figurenum{7}
\epsscale{.85}
\plotone{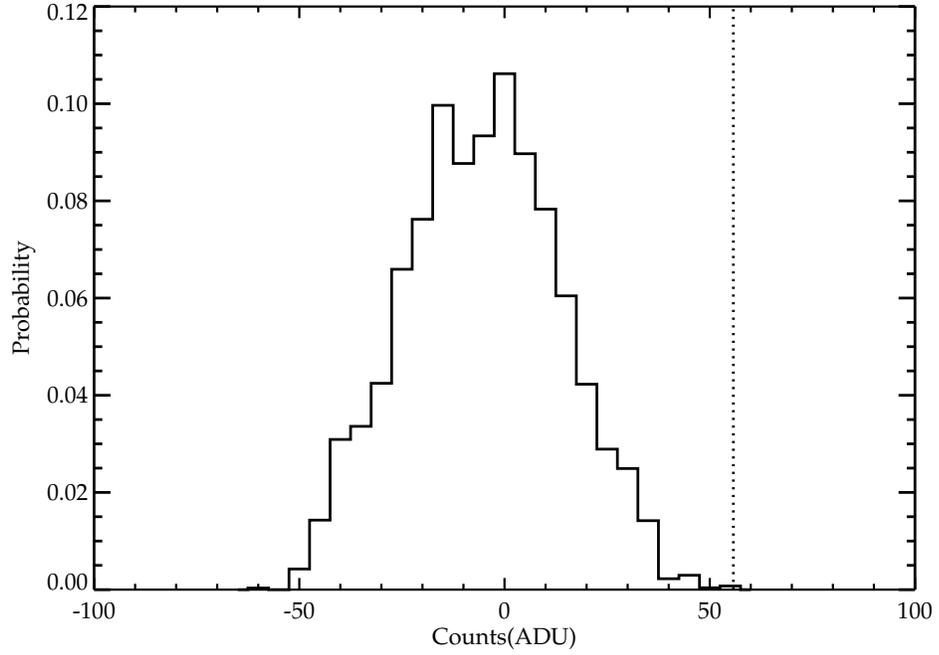}
\caption{Normalized histogram of 
summed flux in randomly placed apertures 
in clean regions of sky-subtracted spectra.  The dotted 
line represents value of a typical candidate Lyman-$\alpha$
emitter.  The frequency of features with fluxes on par
with the candidates is 0.02\%, consistent with a $\simeq$ 
4$\sigma$ detection assuming a Gaussian distribution 
of noise.
}
\label{}
\end{figure}

\begin{figure}
\figurenum{8}
\epsscale{.85}
\plotone{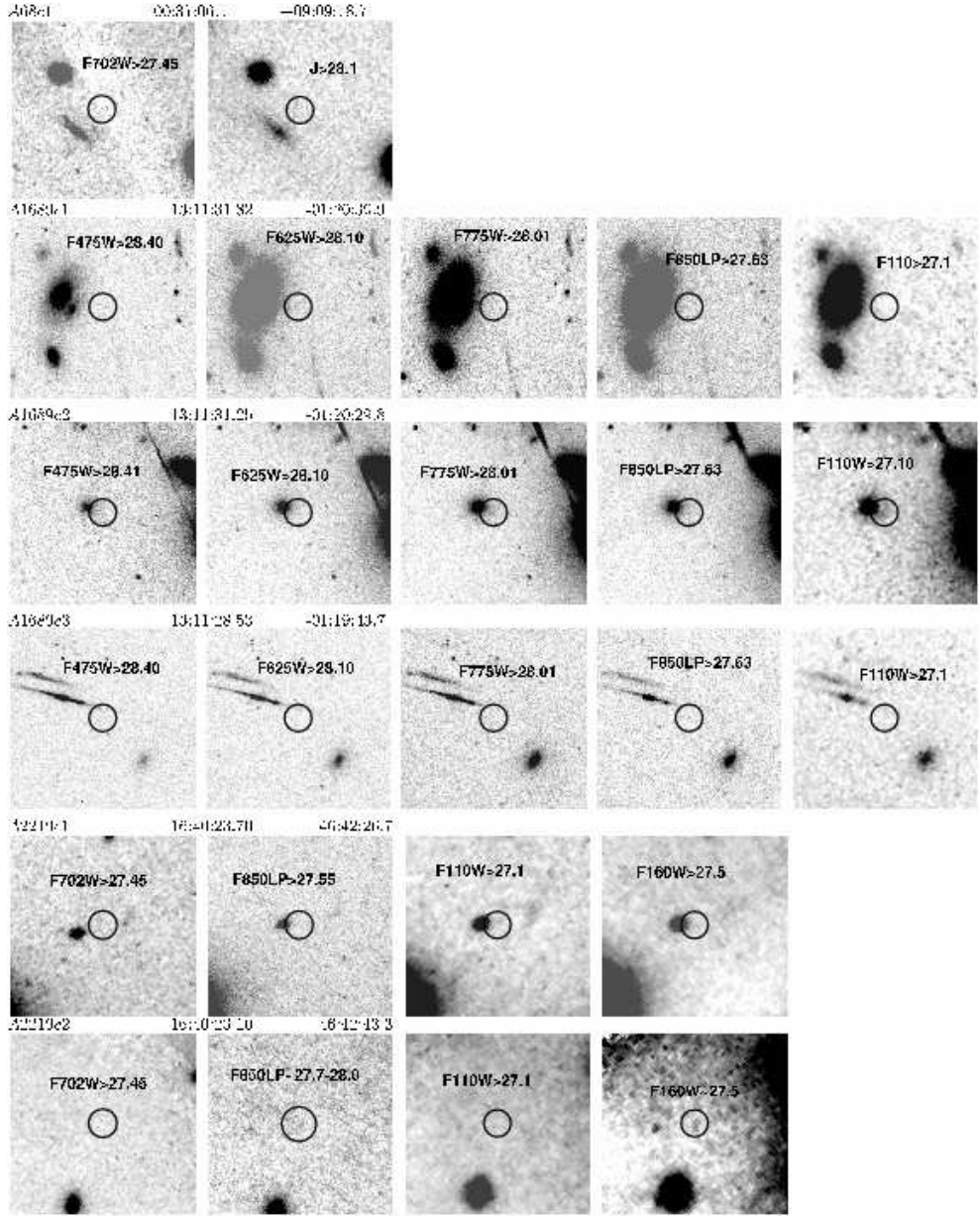}
\caption{Deep broadband images of the locations of the candidate J-band 
emission features.  None of the candidates have {\it definitive} detections 
in the optical; although A2219c2 is very marginally detected in the 
$z'_{F850LP}$ and H$_{F160W}$ bands.  Three of the candidates are located 
within several arcseconds of brighter objects (A2219c1, A1689c1, and A1689c2); however in both 
cases, our registration is known to sufficient precision to rule out 
the possibility that the J-band emission features emanate from the bright
objects. The lack of detection in very deep near-IR images suggests that 
if these candidates are at $z\simeq 9$, then their rest-frame UV continuum 
emission is very faint.  
}
\label{}
\end{figure}

\begin{figure}
\figurenum{9}
\epsscale{.85}
\plotone{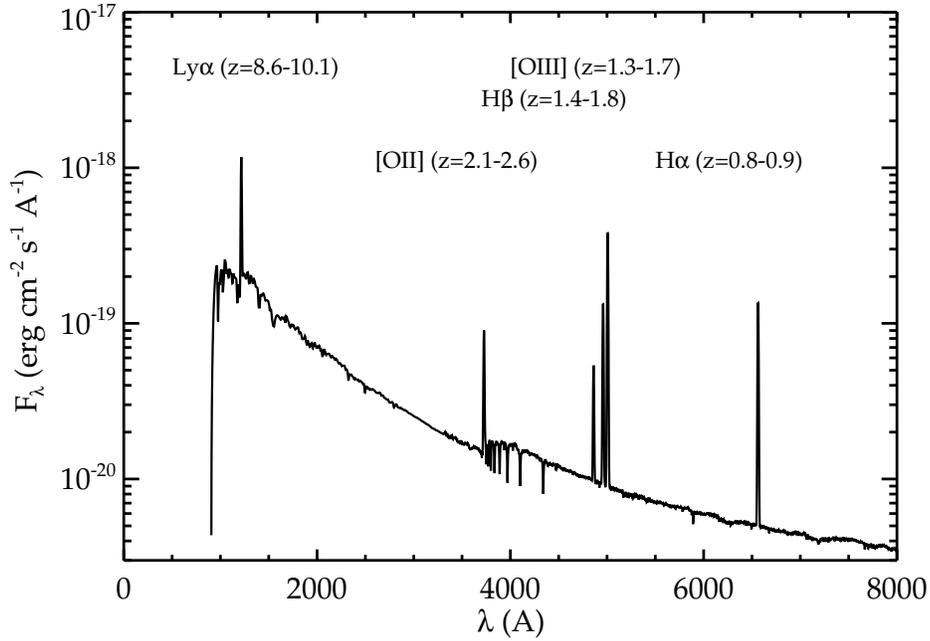}
\caption{Most likely emission features and corresponding redshifts
of the J-band emission features.  If the line lies in the 
rest-frame optical ([OII], H$\beta$, [OIII], H$\alpha$), then additional 
emission features should be present in 
optical or near-infrared spectroscopy. If, on the other hand, the line 
is Lyman-$\alpha$ at $z$=8.5-10, then additional powerful emission 
lines will have been redshifted into the mid-infrared,
making confirmation very difficult.  
}
\label{}
\end{figure}

\begin{figure}
\figurenum{10}
\epsscale{.85}
\plotone{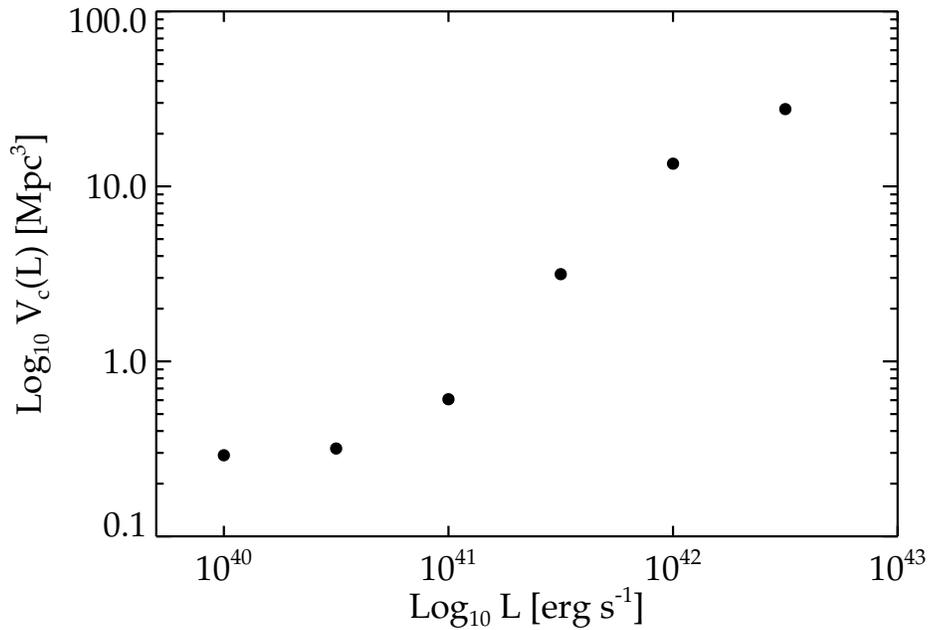}
\caption{Survey volume (comoving) sensitive to sources with Lyman-$\alpha$ 
luminosity L. The volumes observed are very small as a result of the small 
area of the spectroscopic slit and the decrease in survey area resulting 
from gravitational lensing. }
\label{}
\end{figure}

\begin{figure}
\figurenum{11}
\epsscale{.85}
\plotone{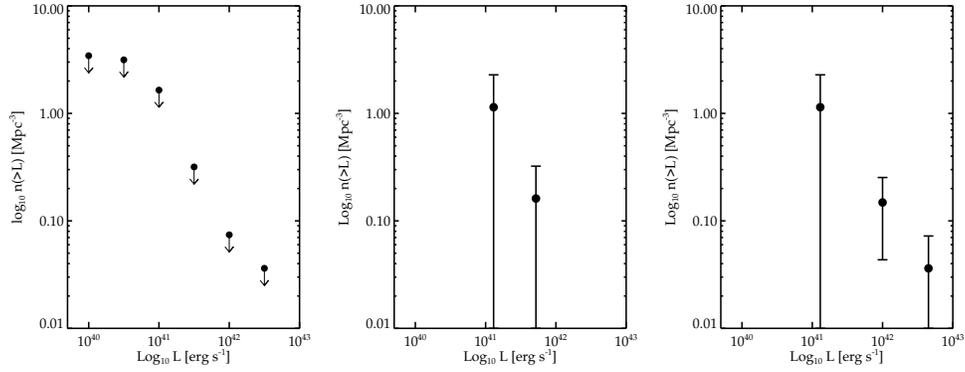}
\caption{Constraints on number density of 8.5$<$z$<$10.2 sources with 
Lyman-$\alpha$ luminosity brighter than L. Left: The 5$\sigma$ 
upper limit to the abundance of low luminosity Lyman-$\alpha$ emitters is 
presented assuming none of our candidate Lyman-$\alpha$ emitters is 
real. Middle: The cumulative abundance of Lyman-$\alpha$ emitters if 
two of the candidates (A68c1 and A2219c1) are real.  Right: The 
cumulative abundance of Lyman-$\alpha$ emitters 
assuming each of the six candidates is a Lyman-$\alpha$ emitter. If 
even one candidate is real, the inferred abundance of low luminosity 
Lyman-$\alpha$ emitters at $z\simeq 9$ would be very high.}
\label{}
\end{figure}

\begin{figure}
\figurenum{12}
\epsscale{.85}
\plotone{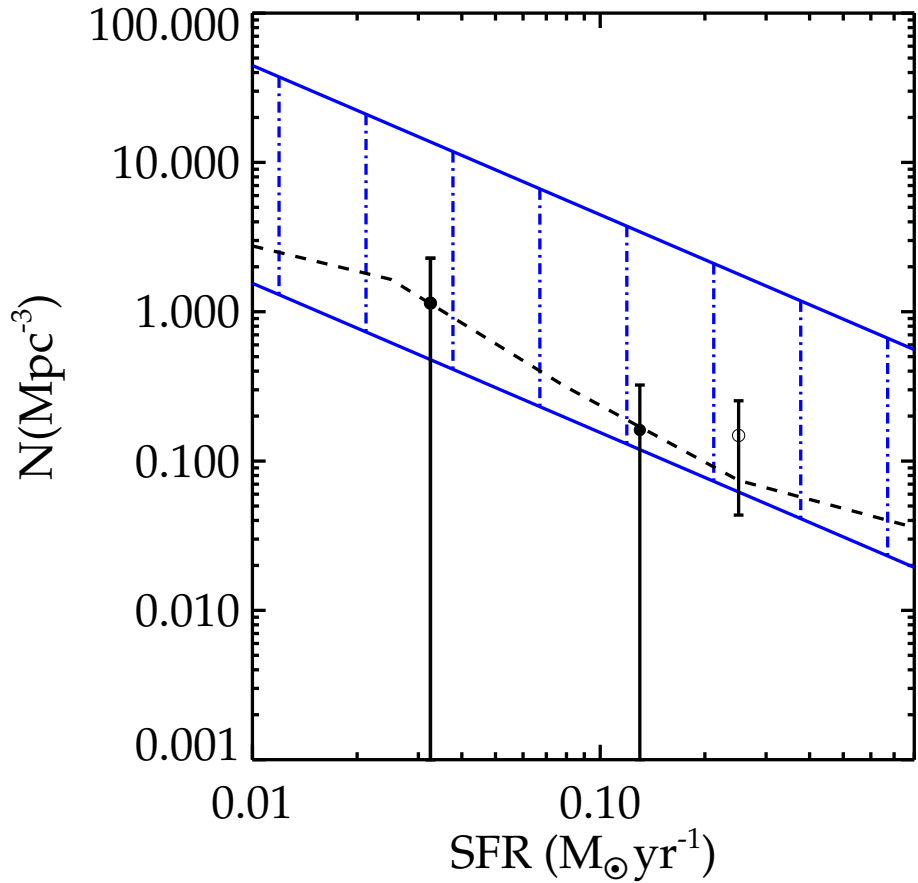}
\caption{Abundance of sources required for reionization assuming
a single population of star-forming galaxies dominates the process 
(blue swath) compared the observed abundances (a) if all the 
candidates are real (open circles), (b) two of the candidates 
are real (solid circles), or (c) if none of the candidates are 
real (dotted line). 
The abundance constraints for case a and case b are identical 
for the two lowest luminosity bins plotted in the figure.
The range for the abundance of sources required for reionization 
is determined assuming f$_c$=0.5, $\Delta t_{reion}$=575 Myr, and B=10 
for the lower locus of points and f$_c$=0.02, $\Delta t_{reion}$=250 Myr, 
and B=5 for the upper locus of points.  If two or more of the 
candidates are real, then it appears that intrinsically faint 
galaxies may play a dominant role in the reionization of neutral
hydrogen.  
}
\label{}
\end{figure}

\end{document}